\documentclass[11pt]{article}

\usepackage{amssymb,amsmath,amsfonts,eurosym,geometry,ulem,color,setspace,sectsty,comment,footmisc,pdflscape,array}
\usepackage[colorlinks=true, linkcolor=blue, citecolor=blue, urlcolor=blue]{hyperref}
\usepackage[capitalize,nameinlink]{cleveref}        
\crefname{app}{Appendix}{Appendices}
\Crefname{app}{Appendix}{Appendices}
\usepackage{fancyhdr} 

\fancypagestyle{plain}{
\fancyhf{}
\fancyfoot[C]{\thepage}
}
\usepackage[style=apa,natbib=true,backend=biber,doi=false,url=false]{biblatex}
\DeclareCiteCommand{\mycite}
  {\usebibmacro{prenote}}
  {\bibhyperref{\printnames{labelname}\space\mkbibparens{\printfield{labelyear}\printfield{extradate}}}}
  {\multicitedelim}
  {\usebibmacro{postnote}}

\usepackage{booktabs}

\usepackage[utf8]{inputenc}
\usepackage{graphicx}
\usepackage{subcaption}
\usepackage{tabularx}
\usepackage{float}
\usepackage{authblk}

\usepackage{enumitem}
\usepackage{multirow}
\usepackage{amsthm}
\newtheorem{assumption}{Assumption}
\crefname  {assumption}{Assumption}{Assumptions} 
\Crefname  {assumption}{Assumption}{Assumptions} 

\newtheorem{theorem}{Theorem}

\crefname{figure}{Figure}{Figures}  
\Crefname{figure}{Figure}{Figures}  

\AddToHook{cmd/appendix/before}{%
  \setcounter{lemma}{0}%
  \setcounter{theorem}{0}%
 }

\newlist{steps}{enumerate}{1}
\setlist[steps, 1]{label = Step \arabic*:}

\newcolumntype{L}[1]{>{\raggedright\let\newline\\arraybackslash\hspace{0pt}}m{#1}}
\newcolumntype{C}[1]{>{\centering\let\newline\\arraybackslash\hspace{0pt}}m{#1}}
\newcolumntype{R}[1]{>{\raggedleft\let\newline\\arraybackslash\hspace{0pt}}m{#1}}

\begin{document}

\begin{titlepage}
\title{Average Treatment Effect Localization: Projection Methods in Synthetic Control}

\author{
Ruei-Chi Lee\thanks{Department of Economics, Rutgers University; email:
\texttt{rl824@rutgers.edu}}}

\date{\today}
\maketitle
\begin{abstract}

 Many real-world policies and business interventions require assessing short-term effects to inform timely decisions, even though most causal inference methods focus on long-term average treatment effects. In this paper, we introduce average treatment effect localization (ATEL), which captures localized, short-term policy impacts in panel data settings with a single treated unit and provides early indicators of policy impact. To accommodate both time-varying and nonlinear effects of observed and unobserved covariates, we propose a nonparametric model for untreated outcome, interpreted as a time-varying factor model via sieve approximation. Estimating the time-varying factor model is challenging due to the boundary bias and identification. Our estimation method based on diversified projection can effectively address these issues. We develop an asymptotic distribution theory to facilitate inference for the ATEL estimator. In an empirical application, we apply our proposed methodology to assess the impact of right-to-carry laws on violent crime rate.\\
\vspace{0in}
\vspace{0in}

\bigskip
\end{abstract}
\setcounter{page}{0}
\thispagestyle{empty}
\end{titlepage}
\pagebreak \newpage

\onehalfspacing

\section{Introduction}

Evaluating policy interventions is crucial, as they are commonly encountered across various fields, including economics, political science, medical science, and others. Empirical researchers often focus on the long-term effects of policy interventions (e.g., \mycite{doi:10.1086/209851, 10.1257/aer.96.3.847, 10.1257/aer.103.6.2052, 10.1257/aer.104.9.2633}). In contrast, short-term policy interventions have received limited attention, partly because short-term effects may lack statistical significance. However, in practice, policymakers and business leaders must assess program effectiveness based on short-term results to inform decision-making and justify continued investment.

In this paper, we estimate the Average Treatment Effect Localization (ATEL) to capture the short-term causal effects localized around the time the treatment is first received. Formally, ATEL is defined as:
\begin{equation*}
    \alpha = \frac{1}{T_1}\sum_{t=T_0+1}^T \mathbb{E}(Y^I_{1t} - Y^N_{1t})K_h(\frac{t-T_0}{T_1})
\end{equation*}
\noindent where $Y^I_{1t}$ and $Y^N_{1t}$ denote the treated and untreated potential outcomes, respectively, for the treated unit at time $t$, and $T_0$ and $T_1$ are the pre-treatment and post-treatment periods. The kernel function $K_h(\cdot)$, with bandwidth parameter $h$, is used to focus on treatment locally near $T_0$. ATEL specifically captures the immediate policy impact occurring soon after the intervention, providing a more precise and practical approach to short-term causal inference.

Although its name resembles the well‑known local average treatment effect (LATE), the two quantities are conceptually distinct. LATE is unit‑local: it averages the causal effect for compliers, the sub‑population whose treatment status changes in response to an instrumental variable \citep{209996d6-9cfc-38cf-a7bb-28a4c5d811aa}. ATEL is time‑local: it averages the causal effect for the treated unit within a narrow window of periods immediately after the intervention.

The relevance of ATEL is particularly evident in contexts such as public policy and corporate strategy. For example, when policymakers implement stricter gun control measures, such as background checks or assault weapon bans, the full impact on crime rates and public safety may take years to materialize. However, legislators and law enforcement agencies often need short-term indicators—such as reductions in firearm-related homicides, gun sales, or illegal firearm trafficking—to evaluate the policy’s effectiveness and justify its continuation. Similarly, in the pharmaceutical industry, when a company launches a new drug, regulatory agencies and investors expect early clinical trial results and real-world patient outcomes before long-term health benefits can be fully observed. Failure to demonstrate early efficacy may lead to funding withdrawals or stricter regulatory scrutiny, even if the long-term benefits are substantial.

This paper studies the problem of making inference and estimation on ATEL in aggregate panel data with a single treated unit. To construct the counterfactual outcomes, we consider the following nonparametric time-varying model for untreated potential outcomes:
\begin{equation}
    Y^N_{it} = h_t(\eta_{it},X_{it}) + u_{it}
\end{equation}
\noindent where $\eta_{it}$ and $X_{it}$ represent unobserved and observed covariates, respectively, $u_{it}$ is the idiosyncratic error, and $h_t(\cdot)$ is a time-varying nonlinear function. This framework allows both unobserved and observed covariates to influence outcomes in nonlinear and time-varying ways, accommodating rich form of heterogeneity than existing methods.

Many existing methods for estimating counterfactuals, such as Difference-in-Differences (DID), rely on strong assumptions like parallel trends, which are often violated due to unobserved time-varying confounders. Synthetic control method (SCM) introduced by \mycite{10.1257/000282803321455188} and \mycite{Abadie01062010} address this by constructing weighted averages of control units to account for time-varying unobservables. Factor-based models (e.g., \mycite{RePEc:tpr:restat:v:98:y:2016:i:3:p:535-551}, \mycite{Xu_2017}) improve on this by using interactive fixed effects, but still impose linear structures on the influence of unobserved confounders. Some studies have introduced nonparametric alternatives (e.g., \mycite{OUYANG2015545}, \mycite{CARVALHO2018352}), but focus on the time-invariant nonlinear function of observed covariates. In contrast, our model provides a general framework that incorporates both observed and unobserved confounders with nonlinear and time-varying effects.

Estimating this model is challenging due to the presence of both unobserved and observed covariates in the time-varying function. We address this issue by assuming the model can be approximated as a low-rank factor structure, interpreted as a sieve approximation: 
\begin{equation}
    h_t(\eta_{it}, X_{it}) \approx \beta_{it}'F_t
\end{equation}

\noindent where $\beta_{it}$ is a vector of sieve basis function $\phi(\eta_{it}, X_{it})$ and $F_t$ is a vector of sieve coefficients.

Since we allow covariates to vary across both time and individuals in the sieve basis function, our model can be approximated as a factor model with time-varying factor loadings. Our work relates to a strand of literature that studies time-varying factor models. For instance, \mycite{SU201784} propose local PCA that combines PCA with the kernel method to estimate time-varying factor models, allowing factor loadings to change smoothly over time. Similarly, \mycite{Pelger03072022} use local PCA with covariate-dependent kernels to estimate time-varying factor models in which covariates vary only across time.

However, estimating time-varying factor models with local PCA has several drawbacks. First, like the kernel regression method, local PCA suffers from boundary bias. Extending it to a local linear framework is infeasible because PCA requires nonnegative kernel weights to maintain the positive semidefiniteness of the covariance matrix, whereas local linear estimators may introduce negative weights to correct boundary bias, particularly near the boundaries. Second, as \mycite{Cheung02012024} points out, while local PCA estimators are only consistent up to a time- or covariate-dependent transformation matrix, $H_t$, they face identification challenges. Specifically, $H_t$ may lack smoothness under some circumstances, resulting in estimation inconsistency. For instance, the order of factors may change over time because PCA sorts factors by eigenvalues, potentially causing abrupt shifts in $H_t$. \mycite{Cheung02012024} provides an example showing that such jumps in $H_t$ can occur over time.

To address the limitations of the local PCA method, we separately estimate the factors and factor loadings. First, to consistently estimate the factors, we adopt a non-PCA-based approach known as diversified projection (DP), proposed by \mycite{doi:10.1080/01621459.2020.1831927} within the pure factor model framework. Unlike PCA, DP provides a more straightforward estimation process that weighted average of observations across exogenous covariates cross-sectionally, such that the transformation matrix, $H_t$, doesn't rely on eigenvalue/eigenvectors. Second, for estimating time-varying factor loadings, we employ the local linear method, which estimates both the factor loadings and their first-order Taylor expansion. This enables the local linear method to effectively mitigate the boundary bias issue inherent in local PCA.

In summary, our contributions can be outlined as follows:
\begin{enumerate}
    \item ATEL is introduced to capture short-term, localized causal effects in panel data with a single treated unit. To facilitate the inference, its asymptotic normality is derived. Unlike conventional methods that focus on long-term average treatment effects, ATEL provides decision-makers with timely indicators of policy intervention impacts.

    \item We propose a general nonparametric model to study treatment effects, which captures the nonlinear and time-varying relationships between observed and unobserved covariates and outcomes. This approach relaxes the restrictive assumptions often imposed on covariates in panel data models for treatment effect estimation.
    \item Our model can be interpreted as a time-varying factor model using sieve approximations. While conventional PCA-based methods face challenges in estimating time-varying factor models, including boundary bias and identification of the transformation matrix, our estimation method effectively addresses these issues.
\end{enumerate}

There is a substantial body of literature studying treatment effects in nonparametric panel data that incorporates both observed and unobserved covariates in the unknown function, known as nonseparable models. Several papers address the time-invariant nonparametric function, including works by \mycite{https://doi.org/10.3982/ECTA8220},
\mycite{HODERLEIN2012300}, and
\mycite{https://doi.org/10.3982/ECTA8405}. More recent studies have developed models where the unknown function varies over time, such as those by \mycite{10.1093/restud/rdx052} and \mycite{10.1093/ectj/utab007}, and \mycite{BOTOSARU2023576}. These studies estimate the partial effects of covariates in the nonparametric function to analyze treatment effects, often requiring identification assumptions such as weak monotonicity or time-invariant structural functions. Our objective, however, is not to estimate the effect of covariates but to control for them. Therefore, our model does not require the identification conditions typically assumed in the nonseparable models.

Our paper is also closely related to the literature on treatment effect estimation under linear factor models; see, for example, \mycite{https://doi.org/10.1002/jae.1230}, \mycite{RePEc:tpr:restat:v:98:y:2016:i:3:p:535-551}, \mycite{Xu_2017},  \mycite{doi:10.1177/00222437221137533}, among others. Recent developments in factor structure causal inference focus on general treatment assignment patterns rather than a single treated unit. \mycite{doi:10.1080/01621459.2021.1967163} propose a two-step PCA estimation for block structure treatment assignments and establish asymptotic analysis for the treatment effects. Other works allow more general treatment/missing patterns by treating the assignment mechanism as random. \mycite{athey2021matrix} considers matrix completion estimator with nuclear norm regularization for causal panel data and derives convergence rates. \mycite{XIONG2023271} apply PCA to a reweighted covariance matrix to accommodate assignment patterns beyond block structures. \mycite{duan2024factor} further extend this framework to non-stationary data and allow the missing probability to depend on fixed effects and factor loadings. \mycite{SU2025106022}propose EM algorithm with nuclear norm regularized estimator as initial value and provide asymptotic theory for the estimators. Different to these methods, we propose to estimate ATEL, a
new estimand that focuses on treatment effects local to the treatment period.

However, these PCA-based approaches rely on a factor structure with time-invariant factor loadings. \mycite{doi:10.1080/01621459.2021.1967163} construct counterfactual outcomes through their “tall-wide” block estimators that depend on the time-invariant transformation matrix. Similarly, the reweighted covariance matrix in \mycite{XIONG2023271} and \mycite{duan2024factor} restricts factor loadings to be time-invariant. Ignoring potential time variation generally leads to inconsistent estimation of the treated unit’s factor loading. It is also possible to handle time-varying loadings in their settings by adopting the local PCA as in \mycite{SU201784}. 
On the other hand, because the DP approach constructs the factor estimator through the cross-sectional average of observations with exogenous covariates, it can be adapted straightforwardly to random or block structure treatment assignment. Nevertheless, as discussed by \mycite{athey2021matrix}, treatment may be given to a single unit (synthetic control) or staggered by experimental design. In such cases, the assumption of random treatment assignment may be inappropriate. Since our analysis focuses on a single treated unit, we leave extensions of the DP approach to random treatment assignment for future research.

The rest of the paper is organized as follows. Section \ref{section2} describes the model and introduces the estimation method for the ATEL estimator. Section \ref{section3} provides the asymptotic analysis
results of the ATEL estimator. In section \ref{section5} we report the simulation results. Section \ref{section6} presents an empirical application that estimates the ATEL of right-to-carry (RTC) laws on violent crime rate.
Section \ref{section7} concludes. All proofs and technical lemmas are provided in the Supplementary Material.

We use the following notation. We denote the maximum and minimum eigenvalues of a square matrix $A$ as $\lambda_{\max}(A)$ and $\lambda_{\min}(A)$. For any matrix $A$, we use $\|A\|$ and $\|A\|_F$ to denote the operator norm and the Frobenius norm respectively. $\lfloor x \rfloor$ denotes the largest integer less than or equal to $x$. For two sequences $a_{NT}$ and $b_{NT}$, we denote $a_{NT} \ll b_{NT}$ (or $b_{NT} \gg a_{NT}$) if $a_{NT} = o(b_{NT})$, $a_{NT} \lesssim b_{NT}$ (or $b_{NT} \gtrsim a_{NT}$) if $a_{NT} = O(b_{NT})$, and $a_{NT} \asymp	b_{NT}$ if $a_{NT} = O(b_{NT})$ and $b_{NT} = O(a_{NT})$
\section{Setup and estimation}\label{section2}
This section provides the main setup for the average treatment effect localization (ATEL) and estimation methods for our model.  
\subsection{The average treatment effect localization }

We consider a panel data setting with $N+1$ units observed over $T$ periods. The treated unit (the first unit) is untreated for $T_0$ period and remains treated thereafter. The $N$ control units remain untreated for all periods. Let $Y^I_{it}$ and $Y^{N}_{it}$ be the potential outcomes of unit $i$ in period $t$ with and without treatment respectively. The individual treatment effect for the $i$th unit at time $t$ is defined as
\begin{equation*}
    \alpha_{it} = Y^{I}_{it} - Y^{N}_{it} 
\end{equation*}

\noindent The fundamental challenge for causal inference is that we cannot observe both outcomes $Y^{I}_{it}$ and $Y^{N}_{it}$ simultaneously. Thus, the observed outcomes $Y_{it}$ are related to potential outcomes as $Y_{it} = D_{it}Y^{I}_{it} + (1 - D_{it})Y^{N}_{it}$, where $D_{it} =1$ if $i=1$ and $t>T_0$ and $D_{it}=0$ otherwise.

In the absence of treatment, outcomes are generated by the following nonparametric model:
\begin{equation} \label{equation1}
\begin{split}
    Y_{it}^N &= h_t(\eta_{it}, X_{it}) + u_{it}, \text{ for } i=1,...,N+1,\text{ } t=1,...,T\\ 
\end{split}
\end{equation}

\noindent where the covariates $(\eta_{it}, X_{it})$ are only partially observed. Specifically, $\eta_{it}$ is a vector of unobserved variables, and $X_{it}$ is a vector of observed covariates. This model captures the nonlinearity and time-varying effects from partially observed covariates $(\eta_{it}, X_{it})$ through the time-varying unknown function $h_t(\cdot,\cdot)$. The term $u_{it}$ represents the idiosyncratic error, which is independent of $(\eta_{it},X_{it})$.

After the treated unit receives a policy intervention or treatment, the treated unit outcome for the post-treatment periods is
\begin{equation} \label{equation2}
Y_{1t} = Y^I_{1t} = h_t(\eta_{1t}, X_{1t}) + \alpha_{1t} + u_{1t},\text{ for } t = T_0+1,\ldots,T
\end{equation}

\noindent Given that the treated unit receives the treatment after $T_0$, the conventional causal analysis often focuses on long-term average treatment effect on the treated. It is formally defined as:
\begin{equation*}
     \alpha_{\text{long-term}} = \frac{1}{T_1}\sum_{t=T_0+1}^T \mathbb{E}(Y^I_{1t} - Y^N_{1t})
\end{equation*}

\noindent where $T_1 = T - T_0$ is the number of post-treatment periods. 

In contrast to long-term average treatment effect, this paper focuses on the policy intervention that is localized at the time of treatment $T_0$, to facilitate the analysis of short-term policy impacts. The parameter of interest is the average treatment effect localization (ATEL), defined as:
\begin{equation}
    \alpha = \frac{1}{T_1}\sum_{t=T_0+1}^T \mathbb{E}(Y^I_{1t} - Y^N_{1t})K_h(\frac{t-T_0}{T_1})
\end{equation}
\noindent where $K_h(\cdot)$ is a kernel function with bandwidth parameter $h$. The kernel function assigns weights to post-treatment periods, localizing the treatment effect around $T_0$, so $\alpha$ represents the short-term effect soon after $T_0$. In particular, if $K_h(\cdot)$ follows a uniform kernel, $\alpha$ simplifies to the average treatment effect over the $\lfloor T_1 h \rfloor$ post-treatment periods. Additional details about the kernel function are provided in \eqref{kernel}.

To motivate why short-term policy evaluation is important, our empirical example examines how right-to-carry (RTC) laws affect violent crime shortly after they are implemented. These laws usually allow people to carry concealed firearms in public with a permit. Supporters believe RTC laws can help citizens protect themselves and deter crime, whereas critics argue that more guns in public increase the risk of violence and accidental shootings. Although many empirical studies have investigated the long-term average treatment effects of RTC laws, the findings remain inconclusive (e.g., \mycite{doi:10.1086/467988}; \mycite{RePEc:oup:amlawe:v:13:y:2011:i:2:p:565-631}; \mycite{https://doi.org/10.1111/jels.12219}). Given that the United States has one of the highest rates of firearm violence among developed countries, it is crucial to evaluate RTC laws promptly.

The short-term policy impact especially matters to policymakers such as governors, legislators, and mayors, who often serve short terms of 2–4 years. If an RTC law is linked to a sudden rise in violent crime soon after it goes into effect, it may cause a political backlash, no matter what the long-term data eventually show. Public perception also plays a large role in decisions, sometimes more so than academic findings. A short-term spike in gun violence could lead to the repeal or tightening of RTC laws. Therefore, policymakers need to pay close attention to the policy’s short-term impact when deciding whether it is effective.

In practice, $\alpha$ represents short-term policy effect right after $T_0$, which is useful for policymakers and business leaders who must make prompt decisions. In contrast, researchers often focus on the long-term average treatment effect on the treated. While $\alpha_{\text{long-term}}$ considers the treatment effect across all post-treatment periods, it can be diminished over time as treated units adapt or respond strategically, making it harder to measure the policy’s direct impact. For example, in February 2025, the Trump administration reduced the USAID budget, forcing multiple labs to shut down and causing an immediate drop in academic output. Over time, academic institutions responded with various mitigation strategies, such as fundraising and layoffs, making it difficult to capture the policy’s original impact.

As reported by \textit{The New York Times}, these cuts at Johns Hopkins University, funded largely by USAID grants for public health and agricultural research, led to the elimination of research projects on tuberculosis, AIDS, and cervical cancer. In response, the university laid off more than 2,000 employees to cope with the reduced budget (\mycite{saul2025federal}). For researchers examining how USAID budget reductions influence academic output, these institutional measures and other external adjustments can distort the long-term average treatment effect, obscuring the policy’s direct influence in a long run. By contrast, our main objective, $\alpha$, measures the policy intervention shortly after $T_0$, which remains largely unaffected by these subsequent actions and countermeasures.

\subsection{Low-rank time-varying factor structure: approximation and estimation}\label{DP section}

In this subsection, we discuss how the nonparametric model \eqref{equation1} can be approximated by a low-rank time-varying factor model with a slowly growing rank, and we outline its estimation methods. 

We motivate the sieve approximation for the nonlinear time-varying function $h_t(\eta_{it},X_{it})$ with two considerations. First, standard nonparametric regression is infeasible due to the presence of unobserved covariates $\eta_{it}$. Unobserved heterogeneity is typically handled via linear transformations (e.g., first-differencing) in linear panel data models to remove unit- or time-invariant components. However, because $\eta_{it}$ varies across both units and time and $h_t(\cdot)$ imposes a nonlinear structure, there is no general rule of transformation to remove $\eta_{it}$. Second, the sieve expansion with truncation allows us to approximately represent the matrix formed by the elements $h_t(\eta_{it},X_{it})$ using a low-rank matrix. This low-rank structure is fundamental to our procedure, as it enables us to recover the relevant low-dimensional components while avoiding fully nonparametric estimation of $h_t(\cdot)$.

Formally, let $\Theta$ be the $(N+1) \times T$ matrix where the $(i,t)$ element of $\Theta$ is $h_t(\eta_{it},X_{it})$. Suppose the unknown function $h_t(\eta_{it},X_{it})$ has the following sieve approximation:
\begin{equation} \label{equation4}
    h_t(\eta_{it},X_{it}) = \sum_{j=1}^J \phi_j(\eta_{it},X_{it}) f_{jt} + r_{it} = \beta'_{it} F_t + r_{it}
\end{equation}

\noindent where $\beta_{it} = (\phi_1(\eta_{it},X_{it}),...,\phi_{J}(\eta_{it},X_{it}) )' \in \mathbb{R}^J$ is a set of sieve transformation of $\eta_{it},X_{it}$ using $\phi_j$ as the basis function (e.g., B-spline, Fourier series, wavelets, polynomial series), $F_t = (f_{1t},...,f_{Jt})'$ is the vector of sieve coefficients, $r_{it}$ is the sieve approximation error. $J$ denotes the number of sieve terms which grow slowly as $N,T \rightarrow \infty$. Let $\Theta_0$ be the $(N+1) \times T$ matrix with elements $\beta_{it}' F_t$, and $R$ be the $(N+1) \times T$ matrix of $r_{it}$. In matrix notation, we represent \eqref{equation4} as:
\begin{equation*} 
\Theta= \Theta_0 + R
\end{equation*} 
\noindent For sufficiently smooth functions $\{h_t(\cdot)\}$, the sieve approximation error $r_{it} \xrightarrow{} 0$ as $J \rightarrow \infty$, so the matrix $\Theta$ can be well approximated by
the low-rank matrix $\Theta_0$. For example, suppose $h_t$ belongs to a H\"older class with parameter $b,C,d > 0$,
\begin{equation*}
    \{ h: \| D^b h(x) - D^b h(y)\| \leq C \|x - y\|^{d}, \text{for all } x,y\}
\end{equation*}
Then, for common bases such as B-splines, trigonometric polynomials, or polynomial series, the approximation error satisfies
\begin{equation*}
    \max_{it} |r_{it}| < C J^{-a},\quad a = (b+d) / (dim(\eta_{it}) + dim(X_{it}))
\end{equation*}
which can be made arbitrarily small for sufficiently smooth functions even if $J$ grow slowly. The smoothness condition justifies the low-rank approximation of $h_t(\eta_{it},X_{it})$ because it guarantees that the approximation error vanishes as $J$ increases, allowing the truncation of the sieve expansion to the first $J$ terms. In many empirical applications, the function $h_t(\cdot)$ represents the average outcome conditional on the characteristics $\eta_{it}$ and $X_{it}$. It is plausible to assume that such function is sufficiently smooth so that it belongs to H\"older class. Then it is well-known that such functions can be well-approximated by sieve functions with a small number of sieve bases, yielding a low-rank structure.

 Combining \eqref{equation1} and \eqref{equation4}, the untreated potential outcomes follow an approximate time-varying factor structure:

\begin{equation} \label{equation5}
    Y^N_{it} = \beta'_{it} F_t + r_{it} + u_{it} 
\end{equation}
Several studies have considered low-rank factor structure to represent nonparametric function using sieve approximation; see, for example, \mycite{10.1093/ectj/utab007}, \mycite{10.1214/23-AOS2293}, and \mycite{CHOI2024105682}. However, instead of time-varying factor structure, these analyses approximate functions as pure factor structure, which restrict the unobserved covariates either be unit-invariant or time-invariant, while reasonable in certain applications, may be unrealistic in many practical settings.

For simplicity, we assume that the rank of the low-rank factor structure, $J$, is known. Since the time-varying function is approximated through a low-rank factor structure via a sieve representation, the rank equals the sieve dimension. When the rank is unknown in an empirical study, consistent estimators of rank 
$J$ can be obtained using a conditioning argument; see, for example, \mycite{https://doi.org/10.1111/1468-0262.00273}, \mycite{https://doi.org/10.3982/ECTA8968}, and \mycite{SU201784}.

Building on the factor representation, we separately discuss the estimation of factors and factor loadings. To estimate factors, we adopt the diversified projection (DP) approach proposed by \mycite{doi:10.1080/01621459.2020.1831927}, which reduces dimensionality by projecting data onto low-dimensional subspaces using a diversified weight matrix $W$. Unlike conventional PCA, which requires a strong signal-to-noise ratio to identify factor estimators, the DP method estimates factors through weighted averages of observations rather than relying on eigenvectors. This projection is conducted across individuals, making the estimator robust to serial conditions. The factor estimator is given by:
\begin{equation} \label{equation6}
    \hat{F}_t = \frac{1}{N} \sum_{i=2}^{N+1} W_{it} Y^N_{it}, \text{ for } t = 1,...,T
\end{equation}

\noindent where $W_{it}=(w_{it1},...,w_{itJ})' \in \mathbb{R}^J$ that should satisfy

\begin{enumerate}[label=(\roman*)]    
    \item $W_{it}$ is independent of $u_{it}$ 
    \item $W_{it}$ is correlated with $\beta_{it}$
\end{enumerate}

Conditions (i) and (ii) are diversified weight's analogous to the exclusion restriction and relevance restriction in instrument variables. They are necessary to consistently estimate $F_t$ up to affine transformation. Substituting \eqref{equation1} and \eqref{equation4} into $\hat{F}_t$ gives
\begin{equation} \label{equation7}
\begin{split}
    \hat{F}_t &= (\frac{1}{N} \sum_{i=2}^{N+1} W_{it} \beta'_{it}) F_t + \frac{1}{N}\sum_{i=2}^{N+1} W_{it} r_{it} + \frac{1}{N} \sum_{i=2}^{N+1} W_{it} u_{it} \\
    & = H_t F_t + e_t
\end{split}
\end{equation}

\noindent where $H_t$ is the $J \times J$ time-varying transformation matrix, $e_t$ is estimation error that includes sieve approximation error $r_{it}$ and idiosyncratic error $u_{it}$. Under the mild conditions of $W$, $e_t \xrightarrow{p}0$ as $N \rightarrow \infty$.

A natural question is how to acquire the
diversified weight matrix that satisfy these two conditions. A compelling choice is through observed characteristics $X_{it}$, which have been extensively explored in factor model. For instance, \mycite{https://doi.org/10.3982/ECTA7432} introduced a characteristic-based factor model to analyze stock returns. This model incorporates factor loadings influenced by firm-specific attributes such as market capitalization and book-to-price ratio. \mycite{fan2016projected} propose a more general form in the factor loading that can be explained by both observed covariates and unobserved components.

We can then define the diversified weight $W_{it}$ as the transformation of the observed characteristics $X_{it}$ with a set of functions $\varphi_j(\cdot):$
\begin{equation*}
    w_{itj} = \varphi_j(X_{it}), \text{ for } i\leq N+1,\text{ }t\leq T, \text{ } j\leq J
\end{equation*}

\noindent where $\varphi_j(\cdot)$ is a sieve basis function such as B-spline, trigonometric series, and polynomial series. Given that the factor loading is expressed as the sieve transformation of covariates (i.e., $\beta_{it} = \phi(\eta_{it},X_{it})$) and $u_{it}$ is independent of $X_{it}$, both conditions required for $W_{it}$ are satisfied. However, certain choices of $W_{it}$ may violate these conditions. For instance, \mycite{doi:10.1080/01621459.2020.1831927} suggest constructing diversified weights from the initial outcome $Y_{i1}$, which has the advantage of not requiring additional observed characteristics. Since $Y_{i1}$ contains the information of $u_{i1}$, if $u_{it}$ is serially correlated, the resulting weights can be correlated with $u_{it}$ and the exogeneity condition fails.

The DP approach used here differs from that in \mycite{doi:10.1080/01621459.2020.1831927} in several ways. First, their DP method is robust to over-estimating the number of factors, it does not require exact knowledge of this number for consistent factor estimation, as long as the rank of the diversified weight, $W_{it}$, is at least as large as the true number of factors. Intuitively, since DP constructs factors as cross-sectional weighted averages rather than eigenvectors, over-estimating the number of factors typically adds only redundant projections onto the same factor space. This differs from PCA, where additional eigenvectors correspond to non-spiked eigenvalues and are often driven by idiosyncratic noise. In contrast, our framework assumes the number of factors, $J$, is known, allowing us to construct a diversified weight matrix with rank equal to $J$. Consequently, our DP method can be viewed as an exactly identified IV approach. It is interesting to extend this framework to overidentified IV settings. We examine robustness to over-estimating the number of factors in the simulation and empirical sections, and leave the associated theoretical development for future work.


Second, \mycite{doi:10.1080/01621459.2020.1831927}'s diversified weights are restricted to being time-invariant, with the recommended choice for $W_i$ based on individual-specific characteristics. Similar restrictions can also be found in some semi-parametric factor models (e.g., \mycite{https://doi.org/10.3982/ECTA7432}; \mycite{fan2016projected}). However, such restrictions may be unrealistic in practice. As discussed in \mycite{https://doi.org/10.3982/ECTA7432}, financial applications often require certain asset characteristics to vary over time. Their work demonstrates the time-series evolution of cross-sectional correlations among asset characteristics, indicating the need for time-variation in asset attributes. Similarly, in macro-level data, characteristics often change over time, as they are aggregated from individual or household-level data. For instance, characteristics such as gender, race/ethnicity, or educational attainment are time-invariant at the individual level but vary over time after aggregation. Thus, the DP approach we consider is more adaptable to practical applications.

For practical implementation of $W_{it}$, since the number of factors, $J$, is equal to the dimension of the diversified weight, we recommend a data-driven rank estimation procedure for choosing $J$. Specifically, we adopt the information criteria suggested by \mycite{SU201784}, which are tailored for time-varying factor structures. For the diversified weight functions $\varphi_j(\cdot)$, we recommend using B-spline bases\footnote{Specifically, for $J=2$ and $J=3$, we use linear and quadratic splines (orders $k=2$ and $k=3$) without knots. For $J \geq 4$ we use cubic B-splines (order $k=4$) with $J-k$ knots, so that the spline space has dimension $J$.}. Although standard sieve bases such as polynomial series, trigonometric series, and B-splines can all generate smooth functions, B-splines offer better numerical stability and more suitable boundary behavior. High-order polynomial series tend to induce severe multicollinearity as the sieve dimension $J$ increases. Trigonometric bases impose a periodic boundary condition on the covariate support, which is typically unnatural for economic covariates. In contrast, B-spline basis functions only affect a local region of the covariate space which provide stable boundary behavior and well-conditioned weight matrices. For an excellent review of the sieve basis function, we refer the readers to \mycite{CHEN20075549}.


\cref{table0} examines the sensitivity of the ATEL estimator to the choice of sieve basis functions, $\varphi_j(\cdot)$, used in the construction of $W_{it}$. We evaluate this across different data-generating processes (DGPs) with the number of factors $J=2$. Specifically, DGP1 specifies $h_t(\cdot)$ as a linear function, DGP2 uses a polynomial-series specification, and DGP3 uses a trigonometric-series specification. More details on these DGPs are provided in \cref{section5}. \cref{table0} reports the simulation coverage probabilities of the 95\% confidence intervals for the standardized ATEL estimates and the mean squared errors (MSEs). Overall, the simulation results are satisfactory. Across all DGPs, empirical coverage is close to the nominal level, and MSEs remain uniformly small. These findings suggest that our estimator is insensitive to the specific choice of sieve basis functions. Additionally, we present further simulations in \cref{section5} across different number of factors $J$ for the construction of $W_{it}$. Those results indicate that our ATEL estimates remain robust to the choice of $J$ as well.

\begin{table}[H]
  \centering
  \begin{tabular}{l c c c c c c}
    \toprule
    & \multicolumn{2}{c}{DGP1} & \multicolumn{2}{c}{DGP2} & \multicolumn{2}{c}{DGP3} \\
    \cmidrule(lr){2-3} \cmidrule(lr){4-5} \cmidrule(lr){6-7}
     & Coverage & MSE & Coverage & MSE & Coverage & MSE \\
    \midrule
    B-spline       & 0.9350 & 0.2023 & 0.9323 & 0.0675 & 0.9400 & 0.0653 \\
    Trigonometric  & 0.9330 & 0.2269 & 0.9300 & 0.0709 & 0.9353 & 0.0691 \\
    Polynomial     & 0.9327 & 0.2055 & 0.9373 & 0.0681 & 0.9407 & 0.0655 \\
    \bottomrule
  \end{tabular}
    \caption{Simulated coverage probability of 95\% confidence interval for the standardized ATEL estimates and MSEs for different sieve basis functions $\varphi_j(\cdot)$ (B-splines, trigonometric series, and polynomial series) under DGP1--DGP3. Notes: the sample size is $N=100$, $T_0 = 50$, $T_1= 50$, and the number of factors $J=2$. Based on simulations with 3000 repetitions.}
    \label{table0}
\end{table}

After estimating the factors, the next step is to estimate the time-varying factor loadings. The conventional PCA may fail under time-varying factor loadings. To see this, let $\tilde{F}_t$ be the PCA estimator of $F_t$. Under regular conditions, the PCA estimator for the factor loading is
\begin{equation*}
\tilde{\beta}_i = \frac{1}{T}\sum_{t=1}^T \tilde{F}_t Y_{it} = \frac{1}{T} \sum_{t=1}^T \tilde{F}_t F_t’ \beta_{it} +o_P(1)
\end{equation*}
\noindent This representation shows that $\tilde{\beta}_i$ is a weighted average (over time) of $\beta_{it}$. As a result, the PCA estimator captures only a global average factor loading rather than the spontaneous, time-varying parameter $\beta_{it}$, and thereby failing to capture the variation in factor loadings over time. 

To estimate time-varying factor loadings, we use local linear regression, a widely used nonparametric method that mitigates boundary bias (see, e.g., \mycite{fan1996local}; \mycite{li2007nonparametric}). By assuming factor loading is a smooth function of $t/T$, we have

\begin{equation*}
    \beta_{it} = \beta_i(\frac{t}{T})
\end{equation*}
Since $\beta_i:[0,1]\rightarrow \mathbb{R}$ is a smooth function of $t/T$, we can approximate $\beta_i(\frac{t}{T})$ by $\beta_i(\frac{r}{T})$ if $r/T$ is close to $t/T$ enough for $r\in \{1,...,T\}$. Specifically, through Taylor expansion, we have
\begin{equation*}
    \beta_i(\frac{t}{T}) = \beta_i(\frac{r}{T}) + \beta_i^{(1)}(\frac{r}{T})(\frac{t-r}{T}) + d_{i}(t,r)
\end{equation*}
\noindent where $\beta^{(1)}_{i}(\cdot)$ is the first-order derivative of $\beta_i(\cdot)$, $d_i(t,r)$ is the Taylor approximation error. 

Based on this local linear approximation, the estimator for the time-varying factor loading is obtained by minimizing the locally weighted sum of squared residuals:
\begin{equation}\label{equation12}
    \min_{\beta_{1r},\beta^{(1)}_{1r}} \frac{1}{T_0} \sum_{t=1}^{T_0} [Y^N_{1t} - \beta_{1r}'\hat{F}_t - \beta^{(1)'}_{1r}\hat{F}_t (\frac{t-r}{T_0}) ]^2 K_h(\frac{t-r}{T_0}),  \text{ for }r = 1,...,T_0
\end{equation}
where $K_h(\cdot)$ is a kernel function and $h=h_{T_0}$ is the bandwidth parameter.
To handle boundary issues and ensure uniform results, we adopt the kernel function as defined in \mycite{SU201784}.

Let $q_{tr}=\Big(\hat{F}'_t,(\frac{t-r}{T_0})\hat{F}'_t\Big)'$ be a $2J\times 1$ vector. Stacking these over $T_0$ periods, we define $Q_r = (q_{1r},...,q_{T_0r})'$ as a $T_0\times 2J$ matrix. Additionally, let $M_r = diag\Big(K_h(\frac{1-r}{T_0}),...,K_h(\frac{T_0-r}{T_0}) \Big)$ be a $T_0\times T_0$ matrix, and $Y_1 = (Y^N_{11},...,Y^N_{1T_0})'$ be a $T_0\times 1$ vector. Then the minimizer of \eqref{equation12} is given by:
\begin{equation}\label{llestimatormatrix}
       \begin{pmatrix}
\hat{\beta}_{1r}\\
\hat{\beta}^{(1)}_{1r} 
\end{pmatrix} = (Q'_rM_rQ_r)^{-1} Q_r'M_rY_1
\end{equation}

\noindent The local linear estimator at the boundary point where $r=T_0$ is used for post-treatment periods. To better illustrate the idea, the Taylor expansion of $\beta_{1t}$ is
\begin{equation*}
    \beta_{1}(\frac{t}{T_1}) = \beta_{1}(\frac{T_0}{T_1}) + \beta_1^{(1)}(\frac{T_0}{T_1})(\frac{t-T_0}{T_1}) + d_1(t,T_0)
\end{equation*}
This motivates our estimator of time-varying factor loading in the post-treatment periods is
\begin{equation}
    \hat{\beta}_{1t} = \hat{\beta}_{1T_0} + \hat{\beta}_{1T_0}^{(1)}(\frac{t-T_0}{T_1}), \text{ for } t = T_0+1,...,T
\end{equation}

 \subsection{Comparison with synthetic control method}  

The synthetic control method (SCM), proposed by \mycite{10.1257/000282803321455188} and \mycite{Abadie01062010}, is  a   widely used approach for estimating the treatment effect. Given the structural similarities in how SCM and our DP approach construct estimators, it is important to compare these methods. SCM also conducts cross-sectional weighted average of untreated outcomes to construct the estimator:
\begin{equation*}
\hat{Y}^{SC}_{1t} =W^{SC'} Y_{t}
\end{equation*}
where $W^{SC} = (W^{SC}_{2},...,W^{SC}_{N+1})$ and $Y_{t} = (Y^N_{2t},...,Y^N_{N+1,t})$ are $N\times 1$ vectors. The weights are selected by best fitting the outcome of the treated unit using pre-treatment data, and the weights are nonnegative and sum to one. Specifically, we focus on the case where all pre-treatment outcomes are included as predictors:
\begin{equation}\label{SCM}
\min_{W \in \Delta } \sum_{t=1}^{T_0}( Y^N_{1t} - W' Y^{N}_{t} )^2
\end{equation}
where $\Delta=\{W \in \mathbb{R}^N : W_i \geq 0 \ \forall i, \sum_{i=2}^{N+1} W_i = 1\}$.

There are two fundamental differences between our DP approach and SCM. First, SCM uses only one weighted vector while we use the $J$ weighting vectors. This difference arises because we use DP to estimate multiple factors, whereas SCM uses a weight to directly estimate the counterfactual outcomes. Second, SCM is primarily motivated by a time-invariant factor structure. To make the comparison transparent, we focus on the case where the untreated outcome follows factor model with time-invariant factor loading with just one factor:
\begin{equation*}
Y_{it}^N = \beta_i F_t+ u_{it},\quad i=1,...,N+1
\end{equation*}
Suppose we just use the SCM weight $W_i^{SC}$ to form our diversified projection:
\begin{equation*}
\hat F_t = \sum_{i=2}^{N+1}W_iY_{it}^N.
\end{equation*}
SCM uses $\hat F_t $ directly as the estimated counterfactual outcome:
\begin{equation*}
\hat F_t  = \hat{Y}^{SC}_{1t}
\end{equation*}
while our DP approach uses $\hat F_t $  as the estimated factor.

 Consequently, our approach requires an additional step to estimate the factor loadings of the treated unit by regressing the treated unit's outcome on the estimated factors over the pre-treatment periods:
\begin{equation*}
\hat\beta_1= (\sum_t \hat F_t^2)^{-1}\sum_t \hat F_tY_{1t}^N
\end{equation*}
\mycite{Abadie01062010} provide conditions under which the existence of $W^{SC}$ requires the factor loadings of the treated unit to be perfectly reconstructed via a convex combination of the factor loadings in the control group; that is,
  \begin{equation}\label{betacondi}
    \beta_1 = \sum_{i=2}^{N+1} W^{SC}_i \beta_i
  \end{equation}
In this case,
\begin{equation*}
\begin{split}
\hat F_t  &=\sum_{i=2}^{N+1}W_i \beta_i F_t+ \sum_{i=2}^{N+1}W_i u_{it}\\
&=\beta_1 F_t+ o_P(1)
\end{split}
\end{equation*}
This shows that under the SCM weight condition,  the DP estimated factors $\hat F_t$ indeed can be used as the counterfactual estimator, which is consistent for $\beta_1 F_t.$

Meanwhile,  we use  $\hat F_t$ as the factor estimator, and apply one further step to estimate $\beta_1$. Under the SCM condition (\ref{betacondi}),  our approach is also consistent. To see this, note that
\begin{equation*}
\begin{aligned}
\hat\beta_1
&=(\sum_t \hat F_t^2)^{-1}\sum_t \hat F_t (F_t\beta_1+u_{1t}) \\
&=(\sum_t \hat F_t^2)^{-1}\sum_t \sum_{i=2}^{N+1}\hat F_tF_t\beta_i W_i
 +(\sum_t \hat F_t^2)^{-1}\sum_t \hat F_t u_{1t} \\
&=(\sum_t \hat F_t^2)^{-1}\sum_t \sum_{i=2}^{N+1}\hat F_t(Y_{it}^N-u_{it})W_i
 +(\sum_t \hat F_t^2)^{-1}\sum_t \hat F_t u_{1t} \\
&=1+\zeta=1+o_P(1)
\end{aligned}
\end{equation*}
where $\zeta = (\sum_t \hat F_t^2)^{-1}\sum_t \hat F_t u_{1t}-(\sum_t \hat F_t^2)^{-1}\sum_t \sum_{i=2}^{N+1}\hat F_tu_{it}W_i.$

As a result,
\begin{equation*}
\hat\beta_1 \hat F_t=  \hat F_t+o_P(1) = \beta_1 F_t+ o_P(1).
\end{equation*}
\noindent Therefore, if the treated unit's factor loading lies within the convex hull of the control group's factor loadings, both methods are consistent. However, when this condition fails, SCM is not consistent. SCM requires the weights to be well-chosen to satisfy the exact matching condition of factor loadings. In contrast, because we flexibly estimate $\beta_1$, our framework does not require the weights to be precisely chosen. Indeed, our requirement for $W$ is weaker than SCM.

Finally, we acknowledge that SCM is simpler to implement which only requires one set of weights. The DP, however,  requires choosing the number of factors and choosing multiple weights to satisfy our conditions.

\subsection{Estimation procedure}
With low-rank approximations, untreated potential outcomes can be modeled as a factor model.
Following the factor-based approach proposed by \mycite{RePEc:tpr:restat:v:98:y:2016:i:3:p:535-551} and \mycite{doi:10.1177/00222437221137533}, we separately estimate the factors and factor loadings (the treated unit). The detailed estimation steps are outlined below:

\begin{steps}
    \item (Initialization) Pre-determine the diversified weights $W_{it}$ as described in Section \ref{DP section}
    \item (Factors $F_t$) Using the DP method, the factor $F_t$ is estimated by projecting the untreated outcomes $Y^N_{it}$ of the control group onto the space defined by the diversified weights $W_{it}$:
    \begin{equation}\label{equation61}
\hat{F}_t=\frac{1}{N}W_t'Y_t, \text{ for } t = 1,...,T
\end{equation}
\noindent where $W_t = (W_{2t},...,W_{N+1,t})’$ is the $N\times J$ matrix.
    \item (Factor loading $\beta_{1t}$) Run local linear estimator $Y^N_{1t}$ on $[\hat{F}_t, \hat{F}_t (\frac{t-T_0}{T_0})]$ with pre-treatment periods, obtain $\hat{\beta}_{1T_0}$ and $\hat{\beta}^{(1)}_{1T_0}$, then
    \begin{equation*}
         \hat{\beta}_{1t} = \hat{\beta}_{1T_0} + \hat{\beta}^{(1)}_{1T_0}(\frac{t-T_0}{T_1}),  \text{ for }t=T_0+1,...,T
    \end{equation*}
    \item (Counterfactual outcome) With the estimated factors and factor loading, the counterfactual outcomes for the treated unit $i=1$ during the post-treatment period $t=T_0+1,...,T$ are constructed as:
    \begin{equation} \label{equation9}
    \hat{Y}^N_{1t} = \hat{\beta}'_{1t} \hat{F}_t, \text{ for } t = T_0+1,...,T
\end{equation}
    \item (ATEL estimator): Using the estimated counterfactual outcomes, the treatment effect for each post-treatment period $(t=T_0+1,...,T)$ is given by:
    \begin{equation}\label{equation20}
        \hat{\alpha}_{1t} = Y_{1t} - \hat{Y}^N_{1t}
    \end{equation}
    The ATEL estimator is then computed as:
    \begin{equation}\label{equation10}
        \hat{\alpha} =  \frac{1}{T_1}  \hat{\alpha}_{1}' \mathbf{K}_{T_1}
    \end{equation}
    \noindent where $\hat{\alpha}_{1} = (\hat{\alpha}_{1,T_0+1},\ldots,\hat{\alpha}_{1T})'$ and $\mathbf{K}_{T_1} = \Big(K_h(\frac{(T_0+1)-T_0}{T_1}),\ldots,K_h(\frac{T-T_0}{T_1})\Big)'$ are $T_1\times 1$ vector.
\end{steps}
Our estimation procedure is computationally efficient and scales well to large panels. It consists of two main steps: (i) factor estimation via DP, a cross-sectional weighted average of untreated outcomes, which can be computationally less costly than PCA because it avoids eigen-decomposition for large matrices; and (ii) factor-loading estimation via local linear smoothing (weighted linear regression), which is a standard procedure and remains easy to compute even if $T_0$ is large, since it uses only the $\lfloor T_0h \rfloor $ observations in a neighborhood of the boundary point $T_0$. In practice, both steps can be implemented efficiently via parallelization over $t$\footnote{To facilitate implementation for applied users, we provide the MATLAB toolbox (\texttt{ATEL}), which is available at \url{https://github.com/rueichilee/ATEL}.}.

\section{Asymptotic results} \label{section3}
\subsection{Inferential theory}
This section presents the inferential theory. We provide the asymptotic normality of the ATEL.
Remind the following notation:
\begin{equation*}
    h_t(\eta_{it}, X_{it}) = \sum_{j=1}^J \phi_j(\eta_{it},X_{it}) f_{jt}+ r_{it} = \beta_{it}'F_t + r_{it}
\end{equation*}

\noindent Let $W_t = (W_{2t},..., W_{N+1,t})'$ be the $N \times J$ matrix. Each of its rows $W_{it}=(w_{it1},...,w_{itJ})'$ is an $J\times 1$ vector for $i=2,...,N+1$ and $t=1,...,T$. $\beta_{t} = (\beta_{2t},...,\beta_{N+1,t})'$ is the $N \times J$ matrix. $F_t = (f_{t1},...,f_{tJ})'$ is a $J\times 1$ vector. $u_t = (u_{1t},...,u_{N+1,t})'$ is $(N+1) \times 1$ vectors for $t = 1,...,T$.

We now state the conditions and asymptotic results of estimators.

\begin{assumption}{(Sieve approximation)}
\label{assumption1}
\begin{enumerate}[label=(\roman*)]

\item $\{h_t(\cdot)  \}_{t\leq T}$ belong to ball $H(U,\|\cdot \|_{L_2},C )$ inside a Hilbert space spanned by the basis $\{\phi_j:j=1,...\}$ with a uniform $L_2-$bound $C$: 
\begin{equation*}
    \sup_{h\in H(U,\|\cdot\|_{L_2})}\|h\|\leq C
\end{equation*}
\noindent where $U$ is the support of $(\eta_{it},X_{it})$
\item The sieve approximation satisfies $\max_{it} |r_{it}|< C J^{-a}$ for some $a,C > 0$.
\item For some $C > 0$, $\max_{j\leq J} \sup_{\eta,X} |\phi_j(\eta,X)|<C$.

\end{enumerate}
\end{assumption}

\cref{assumption1} presents some conditions for sieve approximation. \cref{assumption1} (i) restricts $\{h_t(\cdot)\}_{t\leq T}$ to a Hilbert space with a uniform $L_2$ bound such that $\max_{t\leq T} \sum_{j=1}^\infty f^2_{tj}< \infty$. Moreover, $F_t$ is nonrandom since it's sieve coefficients. Then we illustrate under some sensible conditions on the functional space and the sieve bases, \cref{assumption1} (ii) is well satisfied. Suppose $h_t$ belongs to a H\"older class: For some  $b,C,d > 0$, $\{ h: \| D^b h(x) - D^b h(y)\| \leq C \|x - y\|^{d}, \text{for all } x,y\}$. Moreover, suppose the sieve bases follow some common basis functions such as B-spline, trigonometric polynomials, or polynomial series. Then, with the sufficiently smooth functions $\{h_t(\cdot) \}$, sieve approximation error will satisfy $\max_{it} |r_{it}| < C J^{-a}$, $a = (b+d) / (dim(\eta_{it}) + dim(X_{it}))$. \cref{assumption1} (iii) can be satisfied if the basis functions are bounded such as trigonometric basis or the support of unobserved and observed covariates $\eta_{it}$, $X_{it}$ are bounded. It implies that $\| \beta_{it}\|^2 = O_P(J)$ since factor loading is a vector of sieve basis function and the number of sieve terms grow slowly as $N,T \rightarrow \infty$

\begin{assumption}{(Diversified weights)}\label{assumption2}
There exist a constants $c, C>0$ such that as probability approaching one ($N\rightarrow \infty$),
\begin{enumerate}[label=(\roman*)]
\item $\max_{i,t} |w_{itj}|<C $
\item For each $t$, the $J\times J$ matrix $\frac{1}{N} W_t'W_t$ satisfies \\
\begin{equation*}
    c<\lambda_{\min}(\frac{1}{N}W_t'W_t) < \lambda_{\max}(\frac{1}{N}W_t'W_t  ) <C 
\end{equation*}
\item $W_{it}$ is independent of $u_{it}$

\end{enumerate}
\end{assumption}

\cref{assumption2} ensures that the diversified idiosyncratic error, $\frac{1}{N}\sum_{i=2}^{N+1} W_{it}u_{it}$, satisfies the conditions for a cross-sectional central limit theorem when the idiosyncratic errors ($u_{2t},...,u_{N+1,t}$) are cross-sectionally weakly dependence.  Furthermore, according to \cref{assumption1} (ii), the diversified sieve approximation error $\frac{1}{N}\sum_{i=2}^{N+1}W_{it} r_{it}$ is negligible than the diversified idiosyncratic error such that
\begin{equation*}
    \hat{F}_t - H_tF_t = \frac{1}{N}\sum_{i=2}^{N+1}  W_{it} u_{it} + o_p(1)
\end{equation*}

\noindent Therefore, the $\sqrt{N} (\hat{F}_t - H_tF_t)$ is asymptotically normal when  the idiosyncratic errors ($u_{2t},...,u_{N+1,t}$) are cross-sectionally weakly dependence for each $t \leq T$. In particular, this asymptotic normality result still holds even in the finite $T$, which differs from conventional PCA requiring both $N, T \rightarrow \infty$.

\begin{assumption}{(Factors, factor loadings)}\label{assumption3}
    \begin{enumerate}[label=(\roman*)] 
        \item $\frac{1}{T}\sum_{t=1}^T F_t F'_t \rightarrow \Sigma_F$ for some $J \times J$ positive definite matrix $\Sigma_F$
        \item For each t, there are some constants $c, C>0$ such that as probability approaching one ($N\rightarrow \infty$)
        \begin{equation*}
            c < \lambda_{\min}(\frac{\beta_t'\beta_t}{N})  \leq  \lambda_{\max}(\frac{\beta'_t\beta_t}{N}) < C
        \end{equation*}
    \end{enumerate}
\end{assumption}

\cref{assumption3} impose conditions on factors and factor loadings, similar to those in \mycite{SU201784}. The key difference is that they assume random factors and nonrandom factor loadings, whereas we assume nonrandom factors and random factor loadings, derived from the sieve approximation. \cref{assumption3} (i) ensures the nonsingularity of $\frac{1}{T}\sum_{t=1}^T F_t F_t'$ for sufficiently large $T$. \cref{assumption3} (ii) impose pervasive condition on factor loadings.

\begin{assumption}{(Smoothness over time)}\label{assumption4}

    $\beta_{it} = \beta_i(\frac{t}{T})$ and $W_{it} = W_i(\frac{t}{T})$ are smooth functions with continuous derivatives up to second order on the compact interval $[0,1]$.
\end{assumption}
\cref{assumption4} imposes the common smoothness restriction on time-varying factor loadings typically assumed in the existing literature. The main difference is that, while many studies directly assume factor loadings are functions of $t/T$, we approximate the unknown function $h_t(\eta_{it},X_{it})$ via a sieve approach, where the factor loadings are represented by sieve basis functions of the covariates $\eta_{it}$ and $X_{it}$. The assumption is plausible because aggregate variables often evolve slowly.  In macroeconomic applications, variables such as GDP, inflation, and interest rates are commonly assumed to have trends or low-frequency components that evolve smoothly (see, e.g., \mycite{NELSON1982139}, \mycite{fcb655a2-41a6-360b-b568-962267c5b36b}).

When $\eta_{it}$ and $X_{it}$ follow smooth trajectories over time, i.e., $\eta_{it} = \eta_i(\frac{t}{T})$ and $X_{it} = X_i(\frac{t}{T})$, we have:
\begin{equation*}
    \beta_{it} = \phi(\eta_{it},X_{it}) = \phi(\eta_i(\frac{t}{T}),X_i(\frac{t}{T}) ) = \beta_i(\frac{t}{T})
\end{equation*}

\noindent Likewise, the diversified weight $W_{it}$ is also a sieve basis function of 
$X_{it}$, it follows that $W_{it}$ is a smooth function of time. Accordingly, \cref{assumption4} implies that $H_t = \frac{1}{N}\sum_{i=2}^{N+1}W_{it}\beta_{it}$ is continuous with respect to $t$.

We impose the $\alpha$-mixing condition. Let $\mathcal{F}^0_{-\infty}$ and $\mathcal{F}^{\infty}_T$ denote the $\sigma$-algebra generated by $\{ u_t:t\leq 0\}$ and $\{u_t: t \geq T\}$ respectively. Define the mixing coefficient
$$\alpha(T) = sup_{A\in \mathcal{F}^0_{\infty}, B\in\mathcal{F}^{\infty}_{T} } |P(A)P(B) -P(AB) | $$

\begin{assumption}{(Weakly dependence)}\label{assumption5}
    \begin{enumerate}[label=(\roman*)]
        \item $\{u_t:t\leq T  \}$ is a stationary process, satisfying $E(u_{it}) = 0$ and for some $\delta > 0$, there exists
$C>0$ such that $\mathbb{E}(|u_{it}|^{2+\delta})< C$ for all $i \leq N+1, t\leq T$. Furthermore, $u_{it}$ is independent of $\eta_{it}$ and $X_{it}$ for all $i\leq N+1,t\leq T$
        \item Strong mixing: there exist $C_1 >0$ such that for all $T>0$, $\alpha(T) < exp(-C_1T)$
        \item There is constant $C>0$, so that,\\
        $\max_{j\leq N+1} \sum_{i=2}^{N+1} |\mathbb{E} (u_{it} u_{jt} ) |<C, $\\
        $\max_{t \leq T} \frac{1}{N}\sum_{i,j=2}^{N+1} \sum_{s=1}^T |Cov(u_{it}, u_{js}  )  | < C ,$\\
        $\max_{t\leq T}\max_{i,j\leq N+1} \frac{1}{N} \sum_{q,v=2}^{N+1} \sum_{s=1}^T |Cov(u_{it}u_{qt},u_{js} u_{vs} )|<C $
    \end{enumerate}
\end{assumption}

\cref{assumption5} (i)(ii) is a standard requirement for moments and the mixing coefficient for $\alpha-$mixing time series. In particular, since diversified weight $W_{it}$ is chosen based on $X_{it}$, the independence condition in \cref{assumption5} (i) implies \cref{assumption2} (iii). \cref{assumption5} (iii) allows the idiosyncratic error $u_{it}$ to be weak dependence on both time series and cross-section, which are in the same spirit of approximate factor analysis; see, e.g., \mycite{https://doi.org/10.1111/1468-0262.00273}, \mycite{https://doi.org/10.1111/1468-0262.00392}.

\begin{assumption}{(Signal-to-noise)}\label{assumption6}
\begin{enumerate}[label=(\roman*)] 

      The rank $J$ satisfies:
\item $J^{a} \gg \sqrt{NT_1h}$
\item $\sqrt{N} \gg J$
\end{enumerate}
\end{assumption}

\cref{assumption6} (i) is employed to bound the approximation error of the sieve approximation, requiring $\{h_t(\cdot)\}_{t\leq T}$ to be sufficiently smooth to satisfy this condition.

Local linear regression is applied to estimate $\hat{\beta}_{it}, \hat{\beta}_{it}^{(1)}$. Standard practice involves imposing mild conditions on the kernel functions and the convergence properties of the bandwidth, $h$.

\begin{assumption}{(Kernel estimation)}\label{assumption7}
\begin{enumerate}[label=(\roman*)] 
    \item The kernel function $K^*:[-1,1]\rightarrow \mathbb{R}^+$ is a symmetric bounded probability density function that satisfies Lipschitz continuity. 
    \item As $N,T_0,T_1\rightarrow \infty$, $h\rightarrow 0$, $T_0h \rightarrow \infty$, $T_1h\rightarrow \infty$, $N/T_0h \rightarrow \infty$,  $T_0h^3J^2\rightarrow 0$.
\end{enumerate}
\end{assumption}

\cref{assumption7} is the standard assumption for local linear regression. Several types of kernel function satisfy \cref{assumption7} (i), such as Epanechnikov, uniform and quartic kernels. To handle boundary issues, we define the kernel function $K(\cdot)$ based on $K^*(\cdot)$ following \mycite{SU201784}: 
\begin{equation}\label{kernel}
    \begin{split}
        K_h(\frac{t-r}{T})= \frac{1}{h}K(\frac{t-r}{Th}) = \begin{cases}
      \frac{1}{h} K^*(\frac{t-r}{Th}) / \int^1_{-r/(Th)} K^*(u)du, \quad\text{ if }r\in [1, \lfloor Th \rfloor )\\
      \frac{1}{h} K^*(\frac{t-r}{Th}), \quad\text{ if }r\in [\lfloor Th \rfloor , T-\lfloor Th \rfloor  ]\\
      \frac{1}{h} K^*(\frac{t-r}{Th}) / \int^{(1-r/T)/h}_{-1} K^*(u)du, \quad\text{ if }r\in (T-\lfloor Th \rfloor, T ]\\
    \end{cases} 
    \end{split} 
\end{equation}
To establish an asymptotic result, define $\mu_j=\int u^j K(u)du$ and $\nu_j = \int u^jK^2(u)du$. Specifically, $\mu_0=1$ uniformly.

\vspace{0.5em}

Formally, the following theorem shows that $\hat{\alpha}$ is asymptotic normal:

\begin{theorem}\label{theorem}
    With the \cref{assumption1}-\ref{assumption7}, as $N,T_0,T_1\rightarrow \infty$,
\begin{equation*}
    (\frac{1}{T_0h} \Sigma_1 + \frac{1}{T_1h} \Sigma_2)^{\frac{-1}{2}}(\hat{\alpha} - \alpha) \xrightarrow{d} N(0, 1)
\end{equation*}

\noindent where $\Sigma_1 = \nu_0 \lambda'\xi^{-1}\Omega \xi^{-1}\lambda $, $\lambda = \lim_{T_1 \rightarrow \infty}\frac{1}{T_1}\sum_{t=T_0+1}^T F_t$, $\xi = \lim_{T_0 \rightarrow \infty} \frac{1}{T_0}\sum_{t=1}^{T_0}  F_tF_t' $,\\ $\Omega = \lim_{T_0 \rightarrow \infty} \left[  \frac{1}{T_0}\sum_{t=1}^{T_0} F_tF'_t \mathbb{E}(u^2_{1t}) + \frac{2}{T_0}\sum_{t=1}^{T_0}\sum_{s=t+1}^{T_0}F_tF'_s \mathbb{E}(u_{1t} u_{1s})\right]$,
$\Sigma_2 = \nu_0 \sigma^2$,\\  $\sigma^2 =\lim_{T_1\rightarrow \infty} \left[ \frac{1}{T_1} \sum_{t=T_0+1}^T \mathbb{E}(u_{1t}^2) + \frac{2}{T_1}\sum_{t=T_0+1}^T \sum_{s=t+1}^T \mathbb{E}(u_{1t} u_{1s})\right] $

\end{theorem}

The asymptotic result holds when $N$, $T_0$, and $T_1$ are large, corresponding to the factor-based estimation of the average treatment effect (e.g., \mycite{doi:10.1080/01621459.2021.1967163}, \mycite{doi:10.1177/00222437221137533}). The asymptotic variance comprises two terms. Because the two variance terms vanish at different rate, the rate of convergence of $\hat{\alpha}$ is $1/\delta_{T_0,T_1,h}$ where $\delta_{T_0,T_1,h} = \min\{\sqrt{T_0h},\sqrt{T_1h} \}$. It is important to note that the ATEL estimator is always asymptotic normal regardless of the limit relationship between $T_0$ and $T_1$. To see this, asymptotic distribution can be rewritten as
\begin{equation*}
    \begin{split}
        (\frac{\delta^2_{T_0,T_1,h}}{T_0h}\Sigma_1 + \frac{\delta^2_{T_0,T_1,h}}{T_1h} \Sigma_2  )^{-\frac{1}{2}} (\delta_{T_0,T_1,h} (\hat{\alpha} - \alpha) )  \xrightarrow{d} N(0, 1)
    \end{split}
\end{equation*}

\noindent In particular, if $T_0/T_1\rightarrow 0$, $\sqrt{T_0h}(\hat{\alpha} - \alpha) \xrightarrow{d} N(0, \Sigma_1)$; if $T_1/T_0 \rightarrow 0$, $ \sqrt{T_1h}(\hat{\alpha} - \alpha) \xrightarrow{d} N(0, \Sigma_2)$.

The validity of the asymptotic inference does not require the consistent estimation of $H_t$ because, similar to the standard time-invariant factor model, the rotation matrix is not identifiable. This issue is analogous to the rotation problem in the usual PCA setting, where factors are estimated up to an unknown rotation $H$ that itself cannot be consistently estimated. However, this lack of identification does not affect the validity of asymptotic inference, because the asymptotic variance would depend on $H$ only through either $H F_t$ or $H^{-1} \beta_{i}$. In either case, we can replace these terms with their consistent estimators, $\hat{F}_t$ and $\hat{\beta}_{i}$, to consistently estimate the variance. This analogy carries over to our setting where the rotation matrix $H_t$ varies over time. In other words, the asymptotic variance is rotation-free and can be consistently estimated, even though the rotation matrix $H_t$ itself is not.

We now turn to the consistent estimation of the asymptotic covariance in \cref{theorem}. Given the weak dependence of the time series on the idiosyncratic error, an estimator of $\Sigma_1$ is given by
\begin{equation*}
    \hat{\Sigma}_1 =  \hat{\lambda}'\hat{\xi}^{-1}\hat{\Omega} \hat{\xi}^{-1}\hat{\lambda}
\end{equation*}
\noindent where $\hat{\lambda} = \frac{1}{T_1\mu_0}\sum_{t=T_0+1}^{T} \hat{F}_t K_h(\frac{t-T_0}{T_1})$, $\hat{\xi}= \frac{1}{T_0 \mu_0 } \sum_{t=1}^{T_0} \hat{F_t}\hat{F_t}' K_h(\frac{t-T_0}{T_0})  $, 
\begin{equation*}
    \hat{\Omega} = \frac{h}{T_0} \sum_{t=1}^{T_0} \hat{u}^2_{1t} \hat{F_t} \hat{F_t}' K^2_h(\frac{t-T_0}{T_0}) + \frac{2h}{T_0} \sum_{j=1}^{T_0-1} \sum_{t = j+1}^{T_0} \hat{u}_{1t} \hat{u}_{1j} \hat{F}_t \hat{F}'_{j} K_h(\frac{t-T_0}{T_0})K_h(\frac{j-T_0}{T_0}) 
\end{equation*}

\noindent where $\hat{u}_{1t} = Y_{1t} - \hat{\beta}'_{1t} \hat{F}_t$.

An estimator of $\Sigma_2$ is given by
\begin{equation*}
    \hat{\Sigma}_2 = \frac{h}{T_0} \sum_{t=1}^{T_0} \hat{u}^2_{1t} K^2_h(\frac{t-T_0}{T_0}) + \frac{2h}{T_0} \sum_{j=1}^{T_0-1} \sum_{t=j+1}^{T_0} \hat{u}_{1t} \hat{u}_{1j}K_h(\frac{t-T_0}{T_0})K_h(\frac{j-T_0}{T_0})
\end{equation*}

\noindent If $u_{1t}$ is serially uncorrelated, $\hat{\Sigma}_1$ and $\hat{\Sigma}_2$ can be simplified:
\begin{equation*}
    \hat{\Sigma}_1 = \ \hat{\lambda}'\hat{\xi}^{-1}\hat{V} \hat{\xi}^{-1} \hat{\lambda},\quad \hat{\Sigma}_2 = \frac{h}{T_0} \sum_{t=1}^{T_0} \hat{u}^2_{1t} K^2_h(\frac{t-T_0}{T_0}) 
\end{equation*}

\noindent where $\hat{V} = \frac{h}{T_0} \sum_{t=1}^{T_0} \hat{u}^2_{1t} \hat{F_t} \hat{F_t}' K^2_h(\frac{t-T_0}{T_0})$

In practice, a data-driven approach is recommended for bandwidth selection. One effective method is leave-one-out cross-validation, which determines the optimal bandwidth, $h_{cv}$, by minimizing the following objective function
\begin{equation}\label{hcv}
    h_{cv} = \arg\min_{h} \sum_{t=1}^{T_0}( y_{1t} - \hat{\beta}^{(-t)}_{1t} \hat{F}_t  )^2
\end{equation}

\noindent where $\hat{\beta}^{(-t)}_{1t}$ is the leave-one-out estimator. Specifically, $\hat{\beta}^{(-t)}_{1t}$ is the counterpart of $\hat{\beta}_{1t} = \hat{\beta}_{1T_0} + \hat{\beta}^{(1)}_{1T_0}(\frac{t-T_0}{T_0})$, but excludes the $t-$th observation in the local linear estimation process.

While our main focus is inference for the ATEL, practitioners may also be interested in inference on the estimated counterfactual at each post-treatment period by constructing pointwise confidence intervals. This requires further assumption on $\beta_{it}$. Consider the  estimated counterfactual outcomes at $t>T_0$:
\begin{equation*}
    \begin{split}
        \hat{Y}^N_{1t}  &= -\underbrace{(\hat{\beta}_{1T_0}-H^{-1}_{T_0}\beta_{1T_0})'H_tF_t}_{O_P(\frac{1}{\sqrt{T_0h}})} - \underbrace{(\hat{\beta}^{(1)}_{1T_0}-H_{T_0}^{-1}\beta_{1T_0}^{(1)})'H_tF_t(\frac{t-T_0}{T_1})}_{O_P(\frac{1}{\sqrt{T_0h}})}+\underbrace{\beta'_{1t}F_t}_{O_P(1)}+o_P(1)
    \end{split}
\end{equation*}
The first two terms, estimation error in the factor loading for the pre-treatment periods, are the source of $\Sigma_1$ (before averaging over post-treatment periods), while the third term, $\beta_{1t}'F_t$, is the common components of the factor model. Since we are not considering the treatment effect, the common component $\beta_{1t}'F_t$ is dominant here. If we further assume that $\beta_{1t}$ are i.i.d and follows normal distribution, the statistical inference can still be implemented. For example, suppose $\beta_{1t} \sim N(\bar{\beta}, \Sigma_{\beta} )$ with $\Sigma_{\beta} = diag\Big( \sigma^2_{1},...,\sigma^2_{J}  \Big)$. Then, for each post-treatment period, the variance of the estimated counterfactual outcome is
\begin{equation*}
    \begin{split}
     \tilde{\Sigma}_t = \tilde{\Sigma}_{1,t} + \tilde{\Sigma}_{2,t}
    \end{split}
\end{equation*}
\noindent where $\tilde{\Sigma}_{1,t}=\frac{1}{T_0h}\nu_0 F_t'  \xi^{-1}\Omega\xi^{-1}F_t$ and $\tilde{\Sigma}_{2,t} = F_t' \Sigma_{\beta} F_t$.

The estimator for $\tilde{\Sigma}_{1,t}$ and  $\tilde{\Sigma}_{2,t}$ are given by
\begin{equation*}
    \hat{\tilde{\Sigma}}_{1,t} = \frac{1}{T_0h} \hat{F}_t \hat{\xi}^{-1}\hat{\Omega}\hat{\xi}^{-1}\hat{F}_t, \quad \hat{\tilde{\Sigma}}_{2,t} = \hat{F}'_t \hat{\Sigma}_{\beta}\hat{F}_t
\end{equation*}
\noindent where $\hat{\Sigma}_{\beta} = \frac{1}{T_1}\sum_{t=T_0+1}^{T} ( \hat{\beta}_{1t}-\hat{\bar{\beta}}) ( \hat{\beta}_{1t}-\hat{\bar{\beta}})'K_h(\frac{t-T_0}{T_1})$ and $\hat{\bar{\beta}} = \frac{1}{T_1}\sum_{t=T_0+1}^{T} \hat{\beta}_{1t}K_h(\frac{t-T_0}{T_1})$. Accordingly, the pointwise confidence intervals for the estimated counterfactual outcome can be constructed as
\begin{equation}\label{CI}
    \begin{split}
        \hat{Y}^N_{1t}\pm c_{\alpha} \hat{\tilde{\Sigma}}^{1/2}_t,  \text{ for }t=T_0+1,...,T
    \end{split}
\end{equation}
\noindent where $c_{\alpha}$ is the critical value of a standard normal or $t$-distribution for the given confidence interval, and $\hat{\tilde{\Sigma}}_t$ is the estimator for $\tilde{\Sigma}_t$.

\subsection{Extensions}
We conclude this section with a few extensions of our estimator that highlight the flexibility of our approach.
\subsubsection{Dynamic ATEL}
The original ATEL parameter focuses on the short-term effect localized around the intervention date. However, researchers are often also interested in dynamic treatment effect, that is, how the policy impact evolves after the initial intervention (e.g., \mycite{SUN2021175}, \mycite{CALLAWAY2021200}). In particular, one may ask: How does the effect of treatment vary with the length of exposure? For example, does ATEL increase or decrease as more time elapses after treatment begins?

The ATEL framework can be naturally generalized to a dynamic version that localizes the treatment effect at later post-treatment periods. Specifically, for a given post-treatment horizon $m$, define
\begin{equation*}
\alpha_m = \frac{1}{T_1}\sum_{t=T_0+1}^T \mathbb{E}(Y^I_{1t}-Y^N_{1t}) K_h(\frac{t-(T_0+m)}{T_1})
\end{equation*}
which denotes the dynamic ATEL, capturing the short-term effect localized around the post-treatment $T_0+m$. The corresponding estimator is constructed analogously as
\begin{equation*}
\hat{\alpha}_m =  \frac{1}{T_1}\sum_{t=T_0+1}^T \hat{\alpha}_{1t} K_h(\frac{t-(T_0+m)}{T_1})
\end{equation*}
The asymptotic normality result in Theorem 1 remains valid for $\hat{\alpha}_m$, provided that $m$ is fixed or $m/T_1h \rightarrow 0$.

\subsubsection{ATEL with carryover effects}
In many applications, treatment effects may carry over in time (\mycite{https://doi.org/10.1111/ajps.12685}). Treatment received in earlier periods can continue to influence outcomes in later periods, even if the current treatment status is the same. In other words, the outcome at time $t$ may depend not only on whether the unit is treated at $t$, but also on how long it has been treated and on its past treatment history.

We extend the ATEL framework to accommodate such carryover effects. Let the binary treatment path up to time $t$ be denoted by $D_{1:t}\in \{0,1\}^t$. In the potential outcomes framework, the potential outcome is written as $Y_{1t}(D_{1:t})$, representing the outcome that the treated unit would experience at time $t$ under the treatment path $D_{1:t}$

Following \mycite{https://doi.org/10.1111/ajps.12685}, we impose a limited (finite) carryover assumption that the realization of the potential outcome depends only on the most recent $L$ lags of the treatment path.

\begin{assumption}{(Finite carryover)}\label{assumption8}

For all treatment paths $D_{1:t}\in\{0,1\}^t$,
\begin{equation*}
    Y_{1t}(D_{1:t}) = Y_{1t}(D_{t-L+1},...,D_{t})
\end{equation*}
\end{assumption}
This assumption explicitly restricts carryover effects to at most $L$ periods. Since ATEL measures a local treatment effect using a narrow window in the post-treatment period, it can be generalized to this finite carryover setting. The estimand of interest is now defined as
\begin{equation*} 
\alpha(L) = \frac{1}{T_1}\sum_{t=T_0+L+1}^T \mathbb{E}(Y_{1t}(\boldsymbol{1})-Y_{1t}(\boldsymbol{0})) K_h(\frac{t-(T_0+L)}{T_1}) 
\end{equation*}
\noindent where $\boldsymbol{1}$ and $\boldsymbol{0}$ denote $L$-dimensional vectors of ones and zeros, respectively. Thus $Y_{1t}(\boldsymbol{1})$ and $Y_{1t}(\boldsymbol{0})$ are the potential outcomes for the treated unit that has been continuously treated or untreated over the last 
$L$ periods prior to $t$. 

The key point is that we exclude the first $L$ post-treatment periods, $t=T_0+1,...,T_0+L$, because the corresponding potential outcomes $Y_{1t}(\boldsymbol{1})$ involve a counterfactual history in which the unit would have already been treated for $L$ periods, which is incompatible with the observed treatment path immediately after $T_0$. We therefore construct the ATEL estimator with carryover effects as
\begin{equation*} 
\hat{\alpha}(L) = \frac{1}{T_1}\sum_{t=T_0+L+1}^T \hat{\alpha}_{1t} K_h(\frac{t-(T_0+L)}{T_1}) \end{equation*}
\noindent This estimator is a kernel-weighted average of the post-treatment effects after the policy has been implemented for at least $L$ periods. The asymptotic normality result in Theorem 1 remains valid for $\hat{\alpha}(L)$, since trimming a fixed number $L$ of post-treatment periods is asymptotically negligible.

\section{Simulations}\label{section5}

In this section, we conduct a set of Monte Carlo simulations to evaluate our
theoretical results. We study the asymptotic results in approximating the finite sample distribution of ATEL. To evaluate the performance of our methods, 
we consider three DGPs for the potential outcomes in the absence of the treatment:

\begin{itemize}
    \item DGP1: $Y^N_{it} = X_{it} \theta_t + \lambda_{i1} f_{1t} + \lambda_{i2}f_{2t}+u_{it}$ where $\theta_t \sim N(0,1) \text{ and }\lambda_{i1}, f_{1t},\lambda_{i2}, f_{2t} \sim N(\frac{1}{\sqrt{2}} ,1)$
    \item  DGP2: $Y^N_{it} = h_t(\eta_{it},X_{it}) + u_{it}$ where $h_t(\eta_{it},X_{it}) = \sum_{r=1}^{\infty} \frac{|S_{tr}|}{r^3} \eta_{it}^r + \sum_{r=1}^{\infty} \frac{|V_{tr}|}{r^3} X_{it}^r$
    \item  DGP3: $Y^N_{it} = h_t(\eta_{it},X_{it}) + u_{it}$ where $h_t(\eta_{it},X_{it}) = \sum_{r=1}^{\infty} \frac{|S_{tr}|}{r^3} \sin(r \eta_{it}) + \frac{|V_{tr}|}{r^3} \sin(r X_{it})$
\end{itemize}

\noindent where $X_{it}, \eta_{it}$ are generated from $U(0,1)$, $S_{tr}, V_{tr}$ are generated from $N(2,1)$ and $u_{it}$ is generated from $N(0,1)$. 

Followed from \eqref{equation1} and \eqref{equation2}, the treated potential outcomes are then generated from $Y^I_{1t} = Y^N_{1t} + \alpha_{1t} \text{ for } t = T_0+1,...,T$ with $\alpha_{1t}=0.5$. We consider the several combination of sample size $N= \{30,100,200\}$, $T_0=\{15,60,100\}$, $T_1=\{15,40,100\}$ . Since the rank $J$ is not inferred from the data, we consider $J\in \{2,3,4\}$. 

We construct the diversified weight, $W_{it}$, followed from \cref{DP section}, using observed covariates with B-spline series expansion. For the local linear estimator, the Epanechnikov kernel function $K(u) = \frac{3}{4} (1 - u^2) I(|u|<=1 )$ is used. To select the optimal bandwidth, the leave-one-out cross-validation method is employed.

\cref{table1} reports the simulation coverage probability and average length of $95\%$ confidence intervals for the standardized estimate $\alpha$ based on 3000 simulation replications. Overall, the simulation results are consistent with our asymptotic theory. Across designs, the empirical coverage is close to the nominal level and remains stable to the choices of $J$ in the simulation. In addition, we find that the average length of the confidence intervals under DGP3 is shorter than under DGP1 and DGP2.

\begin{figure}[H]
\captionsetup[subfigure]{labelformat=empty}
  \centering
  \scalebox{0.85}{            
\begin{minipage}{\textwidth}
  \centering

  \begin{subfigure}{0.3\linewidth}
    \centering
    \includegraphics[width=\linewidth]{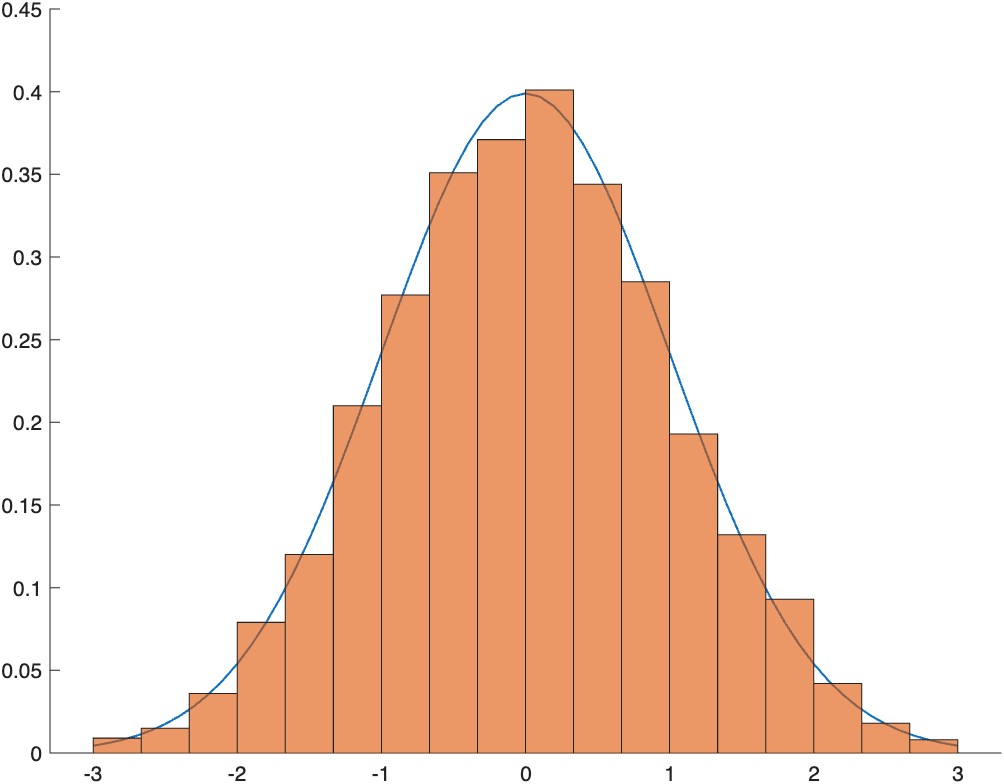}
    \caption{DGP1, J=2}
  \end{subfigure}
  \hfill
  \begin{subfigure}{0.3\linewidth}
    \centering
    \includegraphics[width=\linewidth]{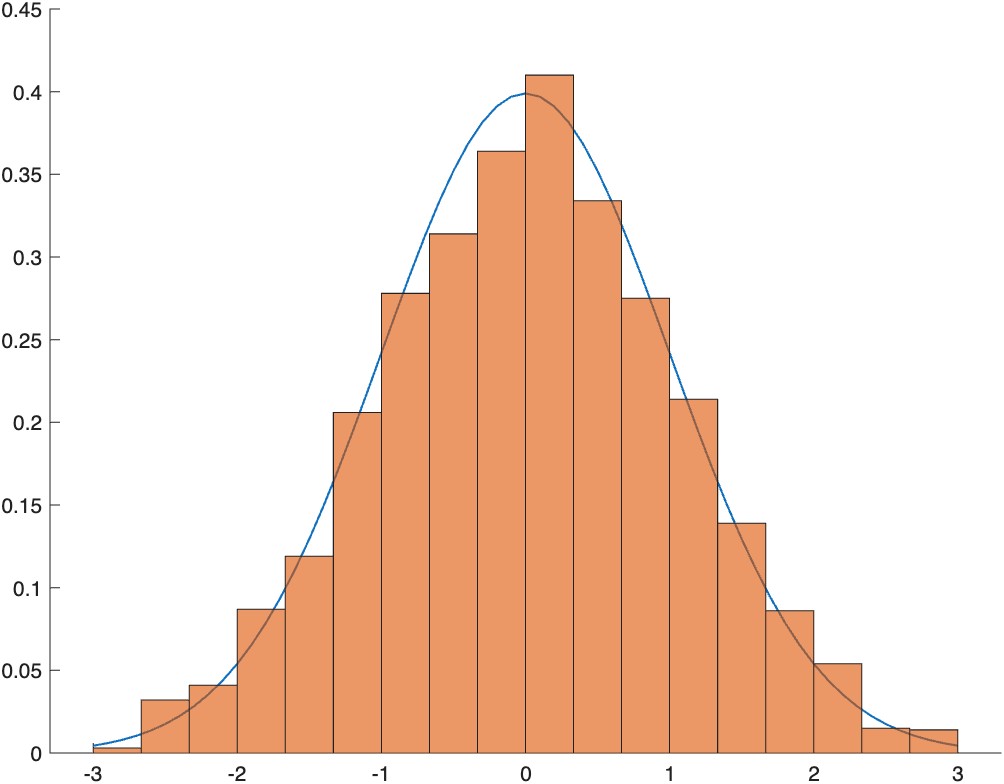}
    \caption{DGP1, J=3}
  \end{subfigure}
  \hfill
  \begin{subfigure}{0.3\linewidth}
    \centering
    \includegraphics[width=\linewidth]{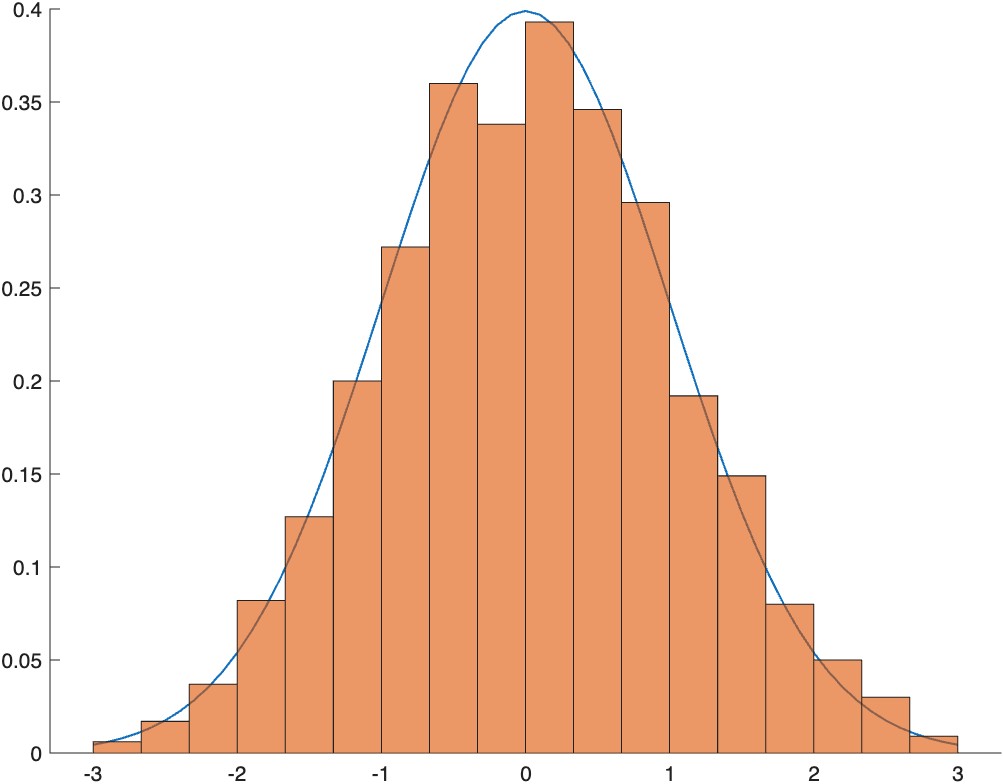}
    \caption{DGP1, J=4}
  \end{subfigure}

  \medskip

  \begin{subfigure}{0.3\linewidth}
    \centering
    \includegraphics[width=\linewidth]{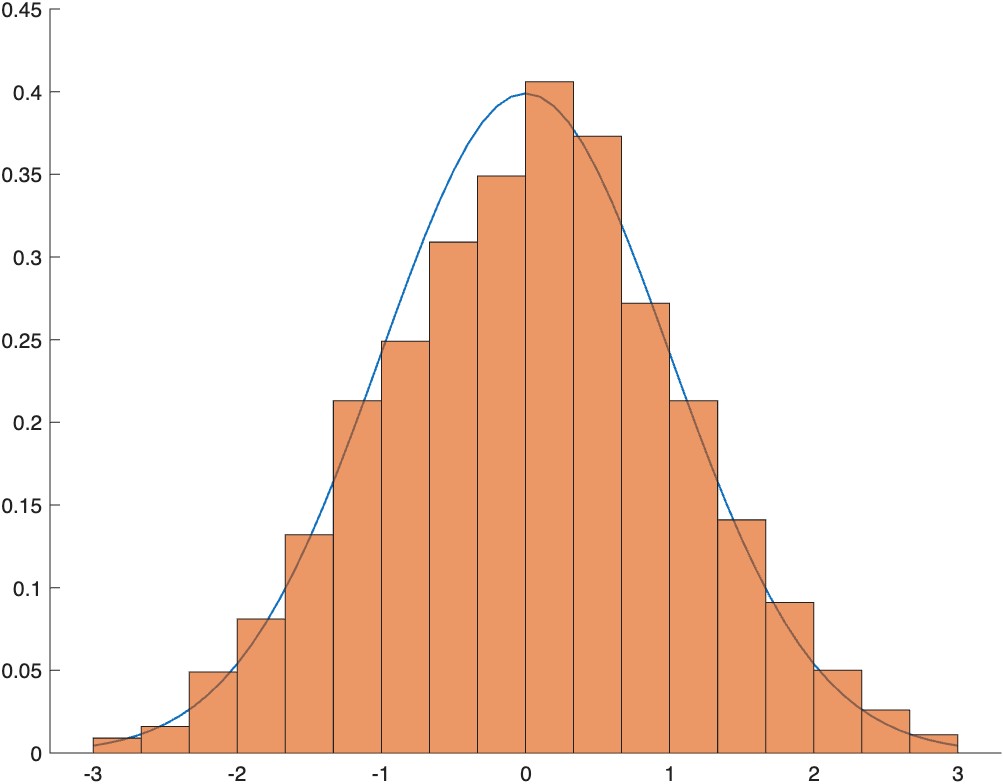}
    \caption{DGP2, J=2}
  \end{subfigure}
  \hfill
  \begin{subfigure}{0.3\linewidth}
    \centering
    \includegraphics[width=\linewidth]{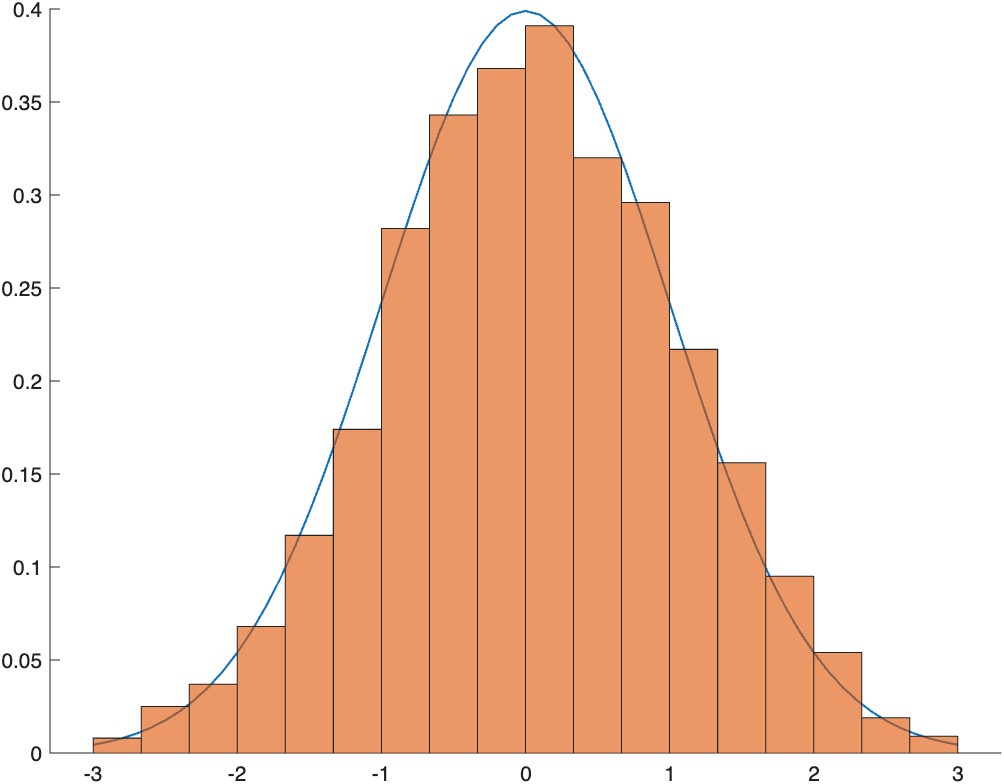}
    \caption{DGP2, J=3}
  \end{subfigure}
  \hfill
  \begin{subfigure}{0.3\linewidth}
    \centering
    \includegraphics[width=\linewidth]{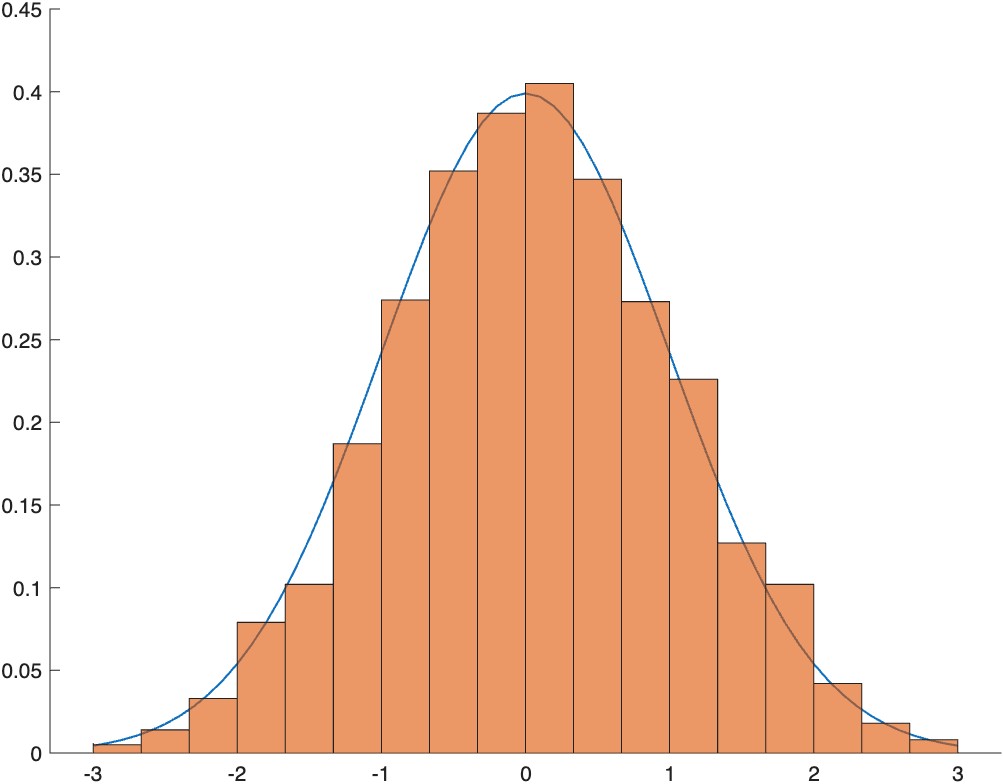}
    \caption{DGP2, J=4}
  \end{subfigure}

  \medskip

  \begin{subfigure}{0.3\linewidth}
    \centering
    \includegraphics[width=\linewidth]{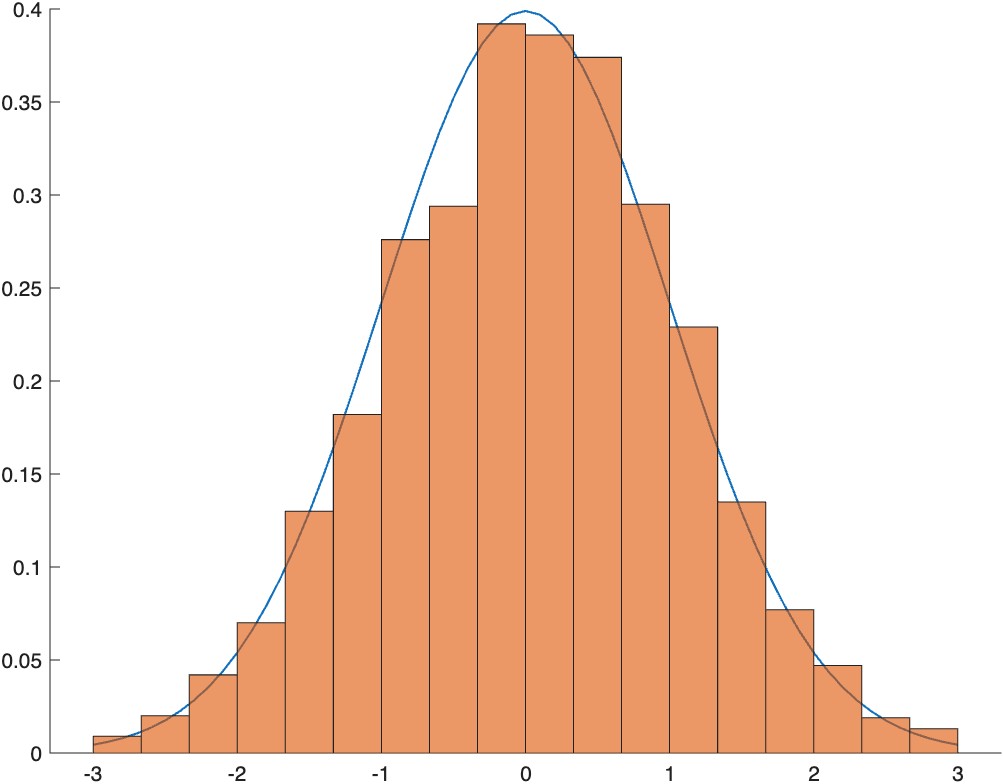}
    \caption{DGP3, J=2}
  \end{subfigure}
  \hfill
  \begin{subfigure}{0.3\linewidth}
    \centering
    \includegraphics[width=\linewidth]{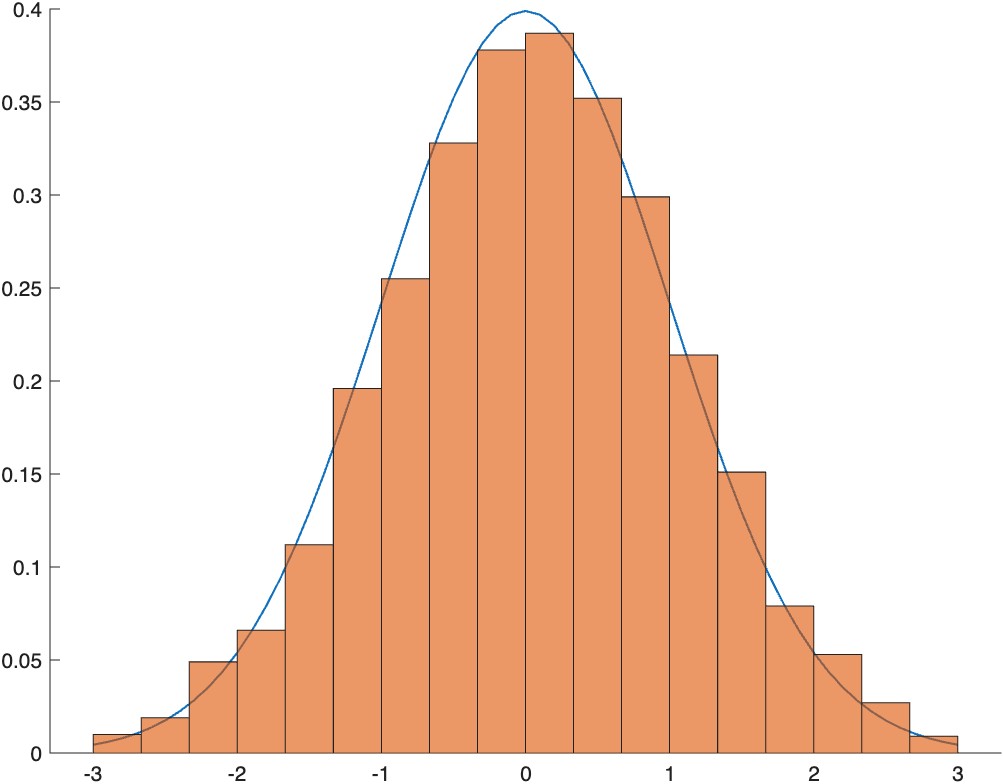}
    \caption{DGP3, J=3}
  \end{subfigure}
  \hfill
  \begin{subfigure}{0.3\linewidth}
    \centering
    \includegraphics[width=\linewidth]{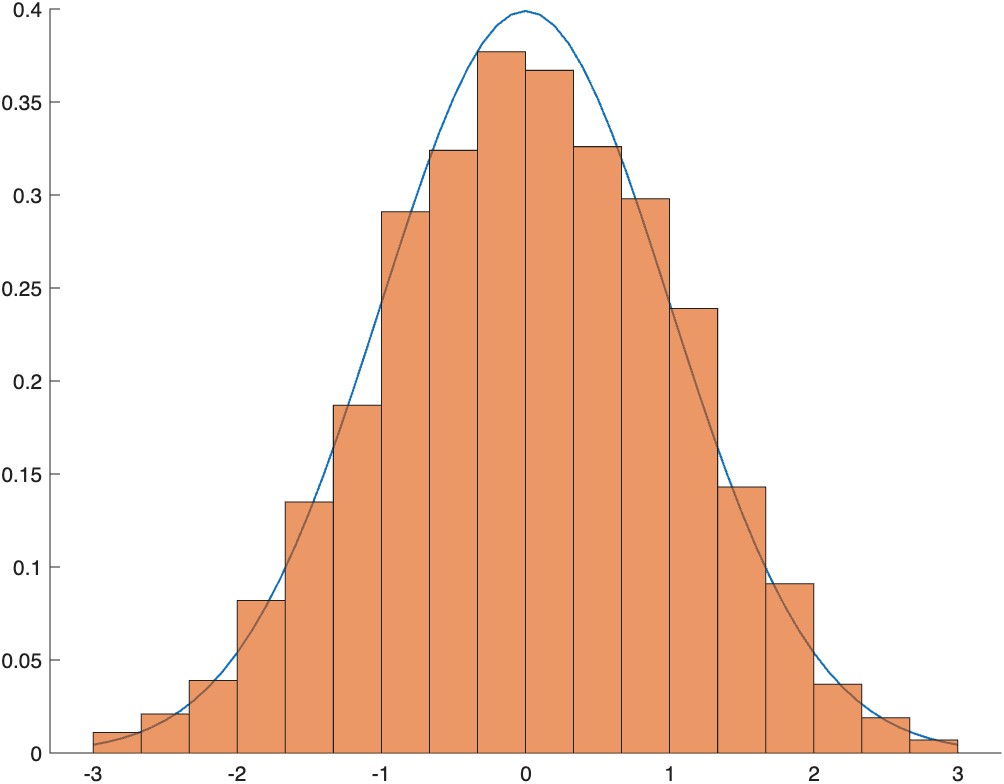}
    \caption{DGP3, J=4}
  \end{subfigure}
  \end{minipage}}
   \caption{Histogram of estimated average treatment effect localization. These histograms are for the standardized estimates $\alpha$ (i.e., $\hat{\alpha} - \alpha $ divided by the estimated asymptotic standard deviation). The standard normal density is superimposed on the histogram. Notes: the sample size is $N=200,T_0=100,T_1=100$. Based on
simulations with 3000 repetitions.}
\label{figure1}
\end{figure}

\cref{figure1} presents the histograms of the standardized estimates of ATEL with the standard normal density. All DGPs exhibit a good approximation to the standardized normal distribution, irrespective of the choice of $J$. This indicates that our inferential procedure is robust to the sieve dimension.

To evaluate the performance of our estimation method, we consider the following DGPs for partially linear models, where the observed covariates $X_{it}$ linearly affect outcome variables and unobserved covariates $\eta_{it}$ in the nonlinear, time-varying function $h_t(\cdot)$. 
\begin{itemize}
    \item  DGP4: $Y^N_{it} = h_t(\eta_{it})+ X_{it}\theta + u_{it}$ where $h_t(\eta_{it}) = \sum_{r=1}^{\infty} \frac{|S_{tr}|}{r^3} \eta_{it}^r$
    \item  DGP5: $Y^N_{it} = h_t(\eta_{it})+X_{it}\theta + u_{it}$ where $h_t(\eta_{it}) = \sum_{r=1}^{\infty} \frac{|S_{tr}|}{r^3} \sin(r \eta_{it}) $
\end{itemize}

\noindent where $X_{it}, \eta_{it}$ is generated from $U(0,1)$, $S_{tr}$ is generated from $N(2,1)$, $\theta$ is generated from $N(\frac{1}{\sqrt{2}},1)$ and $u_{it}$ is generated from $N(0,1)$. 

Since $h_t(\eta_{it})$ is a time-varying function dependent on unobserved covariates $\eta_{it}$, it can also be approximated by a low rank time-varying factor structure, as we did previously, such that $h_t(\eta_{it}) \approx \beta_{it}'F_t$. Therefore, the time-varying factor model can be seen as a special case of model in DGP 4 and 5, where this general nonlinear function is approximated using a time-varying factor structure, and observed covariates affecting the outcome linearly through $\theta$. 

For comparison, in addition to the DP method, we also employ an alternative estimation approach based on the local PCA method of \mycite{SU201784}, who restrict their analysis to a pure time-varying factor structure without observed covariates. In order to estimate $\theta$ simultaneously, we adapt their method by iteratively running local PCA to estimate the time-varying factor structure with observed covariates. This procedure follows the same principle as \mycite{701ec766-57f1-349b-b987-740b9cbdf712}, who propose an interactive fixed-effects model and run PCA iteratively until the coefficient estimates converge. Further details of our estimation scheme are provided in the Supplementary Material.

To estimate the treatment effect using the local PCA method, we proceed as follows:
\begin{steps}
    \item Using local PCA method iteratively, we estimate $F_t$ and $\theta$ using observations in the control group ($i=2,...,N+1$, $t=1,...,T$).
    \item Using local PCA method iteratively, we estimate $\beta_{it}$ using observations in the pre-treatment periods ($i=1,...,N+1$, $t=1,...,T_0$).
    \item Construct counterfactual outcome: $\hat{Y}^N_{1t} = \hat{\beta}'_{1T_0} \hat{F}_t+X_{1t}\hat{\theta}$ and then compute the treatment effect: $\hat{\alpha}_{1t} = Y_{1t} - \hat{Y}^N_{1t}$ for $t=T_0+1,...,T$.
\end{steps}
Because the untreated outcome $Y^N_{1t}$ is unobserved in the post-treatment periods, we use the estimation of $\beta_{1T_0}$ to approximate $\beta_{1t}$ when constructing the counterfactual in Step 3.  To assess the accuracy of the estimated treatment effect, we compute a weighted mean squared error that localizes our estimator around $T_0$ via a kernel function. We refer to this measure as MSEL (mean squared error localization), defined by
\begin{equation*}
    MSEL = \frac{1}{T_1} \sum_{t=T_0+1}^T (\alpha_{1t} - \hat{\alpha}_{1t})^2 K_h(\frac{t-T_0}{T_1})
\end{equation*}
\cref{table2} reports the MSEL results with the number of factor $J=2$. We can observe that our DP method has smaller MSELs compared to the local PCA method across all sample sizes considered. Moreover, DP method provides more stable MSELs as the sample size decreases for both DGP 4 and DGP 5. By contrast, the local PCA method’s MSELs deteriorate with smaller sample sizes. These results indicate that our method performs well even when the model is partially linear.

\begin{table}[H]
\centering

\resizebox{\textwidth}{!}{%
\begin{tabular}{l *{9}{>{\centering\arraybackslash}p{1cm}}}
\toprule
\multicolumn{10}{c}{Coverage}\\
\midrule
 & \multicolumn{3}{c}{DGP1} & \multicolumn{3}{c}{DGP2} & \multicolumn{3}{c}{DGP3} \\
\cmidrule(lr){2-4} \cmidrule(lr){5-7} \cmidrule(lr){8-10}
 & J=2 & J=3 & J=4 & J=2 & J=3 & J=4 & J=2 & J=3 & J=4\\
\midrule
$N=30,T_0=15,T_1=15$ &
{0.9147} & {0.9360} & {0.9417} &
{0.9247} & {0.9340} & {0.9450} &
{0.9217} & {0.9317} & {0.9430} \\
$N=100, T_0=15,T_1=15$ &
{0.9113} & {0.9310} & {0.9357} &
{0.9287} & {0.9297} & {0.9453} &
{0.9223} & {0.9330} & {0.9450} \\
$N=100, T_0=60,T_1=40$ &
{0.9350} & {0.9357} & {0.9440} &
{0.9527} & {0.9357} & {0.9390} &
{0.9360} & {0.9357} & {0.9507} \\
$N=200, T_0=60,T_1=40$ &
{0.9360} & {0.9450} & {0.9447} &
{0.9420} & {0.9320} & {0.9463} &
{0.9453} & {0.9447} & {0.9417} \\
$N=200, T_0=100,T_1=100$ &
{0.9367} & {0.9433} & {0.9423} &
{0.9430} & {0.9397} & {0.9402} &
{0.9487} & {0.9407} & {0.9367} \\
\bottomrule
\end{tabular}%
}

\vspace{0.5cm}

\resizebox{\textwidth}{!}{%
\begin{tabular}{l *{9}{>{\centering\arraybackslash}p{1cm}}}
\toprule
\multicolumn{10}{c}{Average length}\\
\midrule
 & \multicolumn{3}{c}{DGP1} & \multicolumn{3}{c}{DGP2} & \multicolumn{3}{c}{DGP3} \\
\cmidrule(lr){2-4} \cmidrule(lr){5-7} \cmidrule(lr){8-10}
 & J=2 & J=3 & J=4 & J=2 & J=3 & J=4 & J=2 & J=3 & J=4\\
\midrule
$N=30,T_0=15,T_1=15$ &
{3.4371} & {5.7315} & {10.2042} &
{2.0062} & {3.9684} & {5.6613} &
{1.9124} & {2.9438} & {6.8371} \\
$N=100, T_0=15,T_1=15$ &
{3.3668} & {5.8876} & {9.0321} &
{1.9586} & {2.9892} & {5.6640} &
{1.8970} & {2.9984} & {5.6781} \\
$N=100, T_0=60,T_1=40$ &
{1.7794} & {1.7959} & {1.8156} &
{0.9974} & {1.0008} & {1.0319} &
{0.9737} & {0.9785} & {0.9996} \\
$N=200, T_0=60,T_1=40$ &
{1.7995} & {1.7874} & {1.8205} &
{0.9931} & {1.0006} & {1.0208} &
{0.9741} & {0.9818} & {0.9949} \\
$N=200, T_0=100,T_1=100$ &
{1.2271} & {1.2090} & {1.2196} &
{0.6833} & {0.6753} & {0.6804} &
{0.6730} & {0.6661} & {0.6719} \\
\bottomrule
\end{tabular}%
}

\caption{Simulated coverage probability of 95\% confidence interval, and average length CI for the standardized estimate $\alpha$. Based on simulations with 3000 repetitions.}
\label{table1}
\end{table}

\begin{table}[H]
\centering
\setlength{\tabcolsep}{4pt}          
\newcolumntype{C}{>{\centering\arraybackslash}p{2cm}}  

\begin{tabular}{l*{4}{C}}            
\midrule
      & \multicolumn{2}{c}{DGP4} & \multicolumn{2}{c}{DGP5} \\
\cmidrule(lr){2-3} \cmidrule(lr){4-5}
      & DP & local PCA & DP & local PCA \\
\midrule
$N=30,T_0=15,T_1=15$      &    0.9655 &   2.3463 &  0.9577 &   2.3093 \\
$N=100, T_0=15,T_1=15$    &  0.9680 &  2.1321 &   1.0361 &   2.1156 \\
$N=100, T_0=60,T_1=40$    &  1.3245 &   1.4523 &  1.2871 &   1.3776 \\
$N=200, T_0=60,T_1=40$    &  1.2979 &  1.3966 &   1.2667 &  1.3301 \\
$N=200, T_0=100,T_1=100$  &  1.3024 &   1.3674 &   1.2451 & 1.2977 \\
\midrule
\end{tabular}
\caption{Comparison of MSEL between DP and local PCA with $J=2$. Based on simulations with 1000 repetitions.}
\label{table2}
\end{table}

\section{Empirical Application: ATEL of right-to-carry laws on
violent crime}\label{section6}

We illustrate the usefulness of the DP inference with an application about the short-term effects of right-to-carry (RTC) law on violent crime. Determining whether a policy achieves its intended effect in a timely manner is crucial for making future policy decisions, especially in the realm of firearm regulation. Among developed countries, the United States has the highest rate of firearm violence, with gun homicides occurring at a rate roughly 25\% higher than in other high-income countries (Bangalore and Messerli, 2013). Over the past several decades, lawmakers have pursued various strategies to address gun violence by regulating the purchase, possession, and use of firearms. Although the long-term average treatment effects of firearm policies have been studied extensively, relatively little attention has been given to their short-term impacts.

RTC laws, which allow individuals to carry concealed firearms in public, aim to regulate the possession and use of concealed firearms, balancing personal safety with public security. The primary debate of RTC laws revolves around their impact on public safety. On one hand, these laws could reduce violent crime by deterring potential offenders who fear encountering armed victims. On the other hand, they could increase violent crime if permit holders engage in criminal activities or if the presence of concealed firearms exacerbates criminal behavior among others.

Empirical research on RTC laws has struggled to find consensus. For example, \mycite{doi:10.1086/467988} and \mycite{latulip2000more} propose the “More Gun, Less Crime” hypothesis, using panel data analysis of county-level data to support their theory that RTC laws reduce crime rate in every violent crime category between 5 and 8 percent. However, more recent studies, employing updated data and different statistical methods, suggest that RTC laws have minimal impact or may increase violent crime rates (e.g. \mycite{Siegel2019TheIO}, \mycite{doi:10.2105/AJPH.2019.305307}, \mycite{NBERw30190}). For instance, \mycite{RePEc:oup:amlawe:v:13:y:2011:i:2:p:565-631} found that RTC laws have negligible effects on murder rates but increase other violent crime rates by 20 to 30 percent. Similarly, \mycite{https://doi.org/10.1111/jels.12219}, using synthetic control methods, observed that RTC laws are associated with higher violent crime rates after adoption.

We investigate the impact of RTC law implementation in Arizona, where RTC legislation took effect in 1994. Prior to 1994, Arizona prohibited concealed firearms for the general public, categorizing it as a may-issue state, with only law enforcement personnel or those with strong ties to law enforcement allowed to carry concealed firearms. In 1994, Arizona transitioned to a shall-issue state by passing legislation allowing state-permitted gun owners to carry concealed weapons.

We use state-level data on annual violent crime rates (per 100,000 residents) from 1977 to 2006 from the FBI's Uniform Crime Reports (UCR)\footnote{The data is avaiable from \url{http://works.bepress.com/john_donohue/89/}  }. The analysis distinguishes between the pre-treatment period (1977–1993) and the post-treatment period (1994–2006). The potential control group comprises 14 states: 9 states with no RTC legislation before 2014 and 5 states that enacted RTC laws after 2006, such as Nebraska and Kansas. The explanatory variables include the police rate, and the poverty rate. Using these explanatory variables, we construct diversified weights based on a $J$-dimensional B-spline series expansion.

\begin{figure}[H]
    \centering
    \scalebox{1}{
    \begin{minipage}{\textwidth}
        \centering
        \begin{subfigure}[b]{0.45\textwidth}
            \centering
            \includegraphics[width=\linewidth]{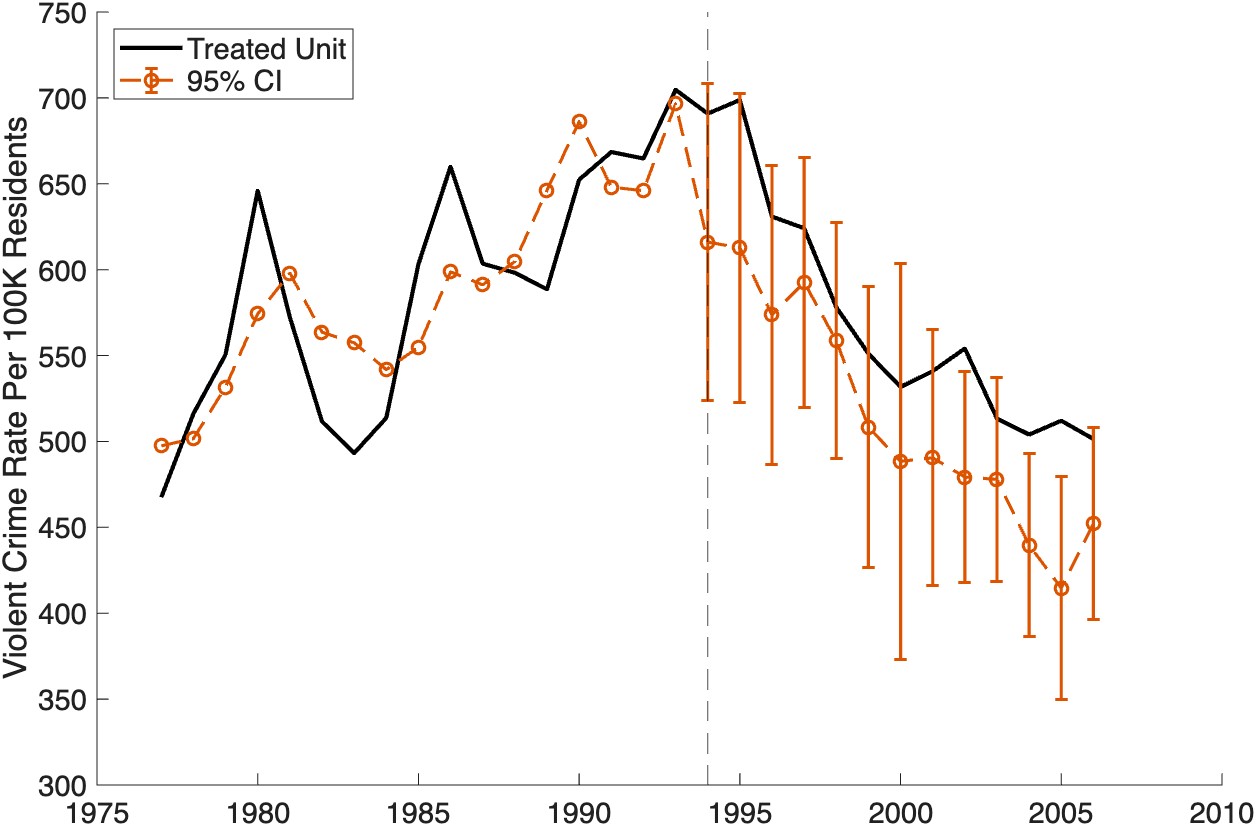}
            \caption{J=2}
        \end{subfigure}
        \hfill
        \begin{subfigure}[b]{0.45\textwidth}
            \centering
            \includegraphics[width=\linewidth]{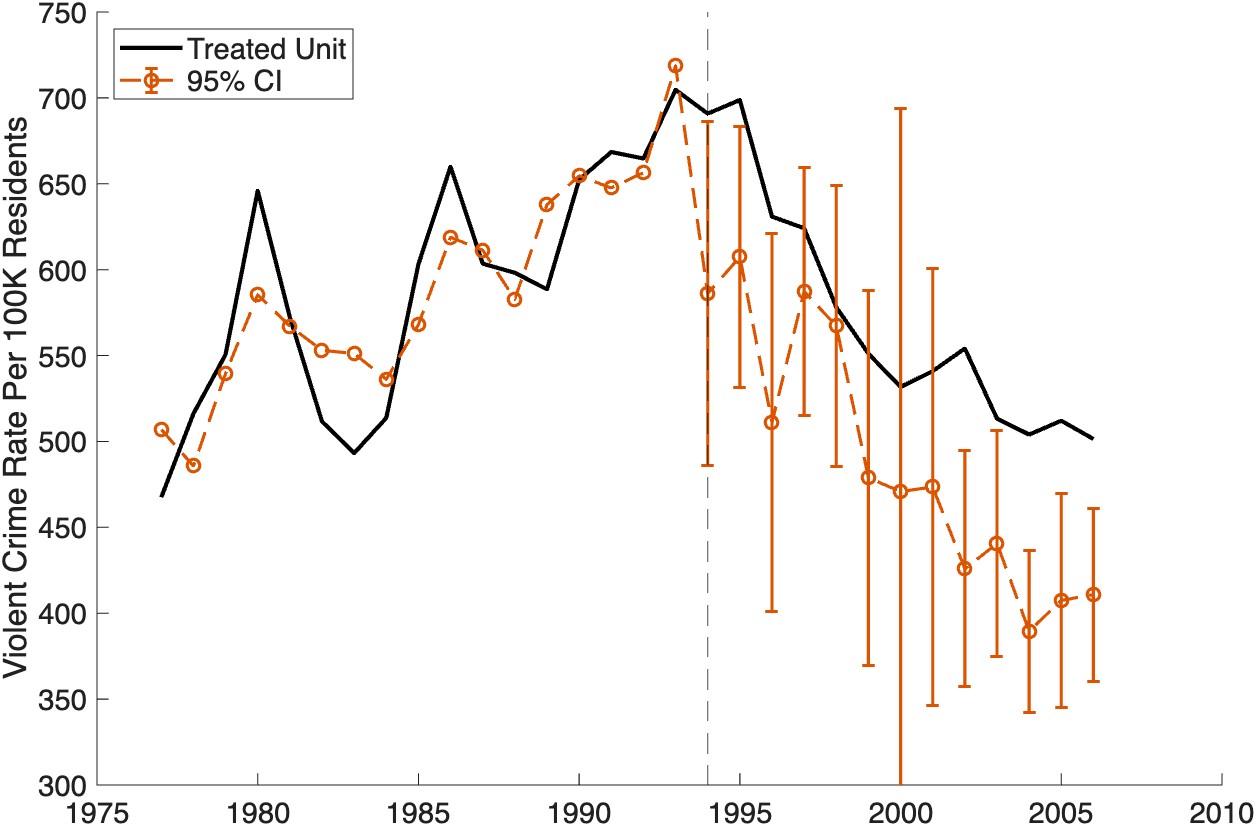}
            \caption{J=3}
        \end{subfigure}

        \vspace{0.5em}

        \begin{subfigure}[b]{0.45\textwidth}
            \centering
            \includegraphics[width=\linewidth]{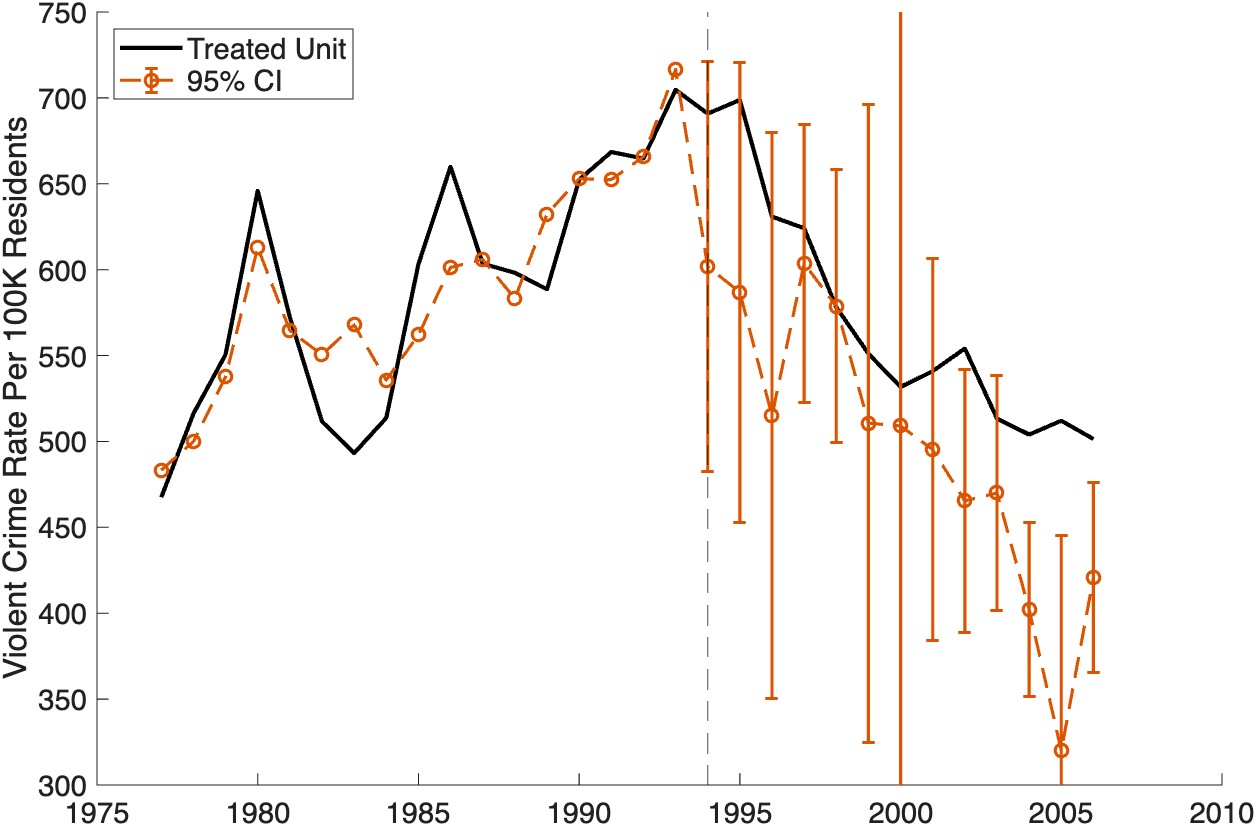}
            \caption{J=4}
        \end{subfigure}
        \hfill
        \begin{subfigure}[b]{0.45\textwidth}
            \centering
            \includegraphics[width=\linewidth]{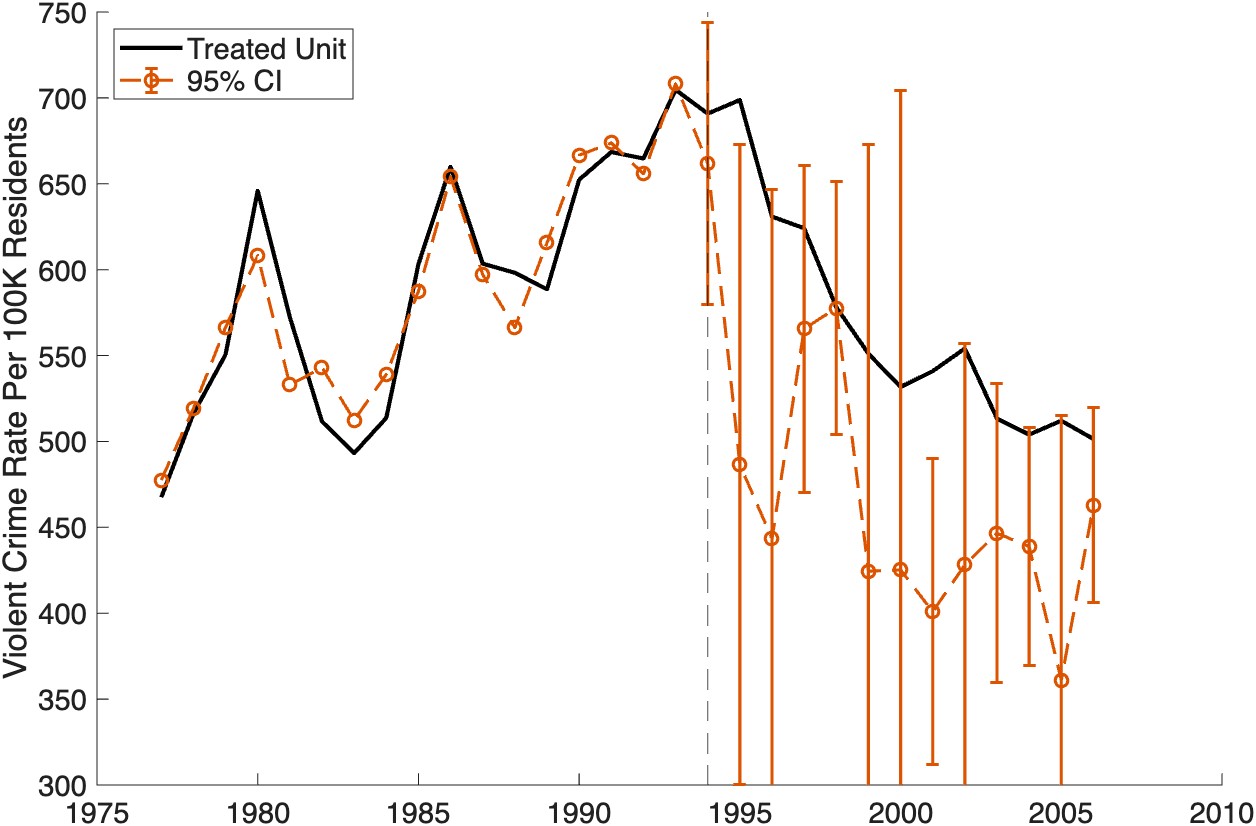}
            \caption{J=5}
        \end{subfigure}
    \end{minipage}
    }

    \caption{Treatment effect of the RTC law on Arizona with different $J$. 
    The solid line is the actual violent crime rate. 
    The dashed lines are the counterfactual outcomes with different $J$. The vertical lines are the pointwise 95\% confidence intervals. The vertical dashed line denotes the beginning of the treatment.}
    \label{figure2}
\end{figure}

Using the data-driven factor selection procedure of \mycite{SU201784}, the information criteria select $J=2$. As a robustness check, we also consider alternative specifications with $J = 2,3,4,$ and $5$. The ATEL estimation results are presented in \cref{figure2}, which displays the actual violent crime rate and estimated counterfactual outcomes during the period 1977-2006. We also report the corresponding pointwise 95\% confidence intervals for the post-treatment periods, constructed under the normality assumption on the factor loadings in \eqref{CI}. We observe that before the intervention occurs, the fitted counterfactual lines closely follow the actual violent crime rate, indicating a good in-sample fit. After the intervention, the counterfactual crime rates decline more sharply than the actual rates, suggesting a positive treatment effect of the RTC law. This pattern holds across all values of $J$. The numerical analysis in \cref{table3} reports the ATEL estimates for different values of J. We can observe a consistent result that the ATEL is statistically significant at the 5\% level for all choices of $J$. These results demonstrate that over-estimating the number of factors maintains satisfactory estimation accuracy, suggesting that our method does not rely on accurate estimation of $J$.

\begin{figure}[H]
    \centering
    \includegraphics[scale = 0.22]{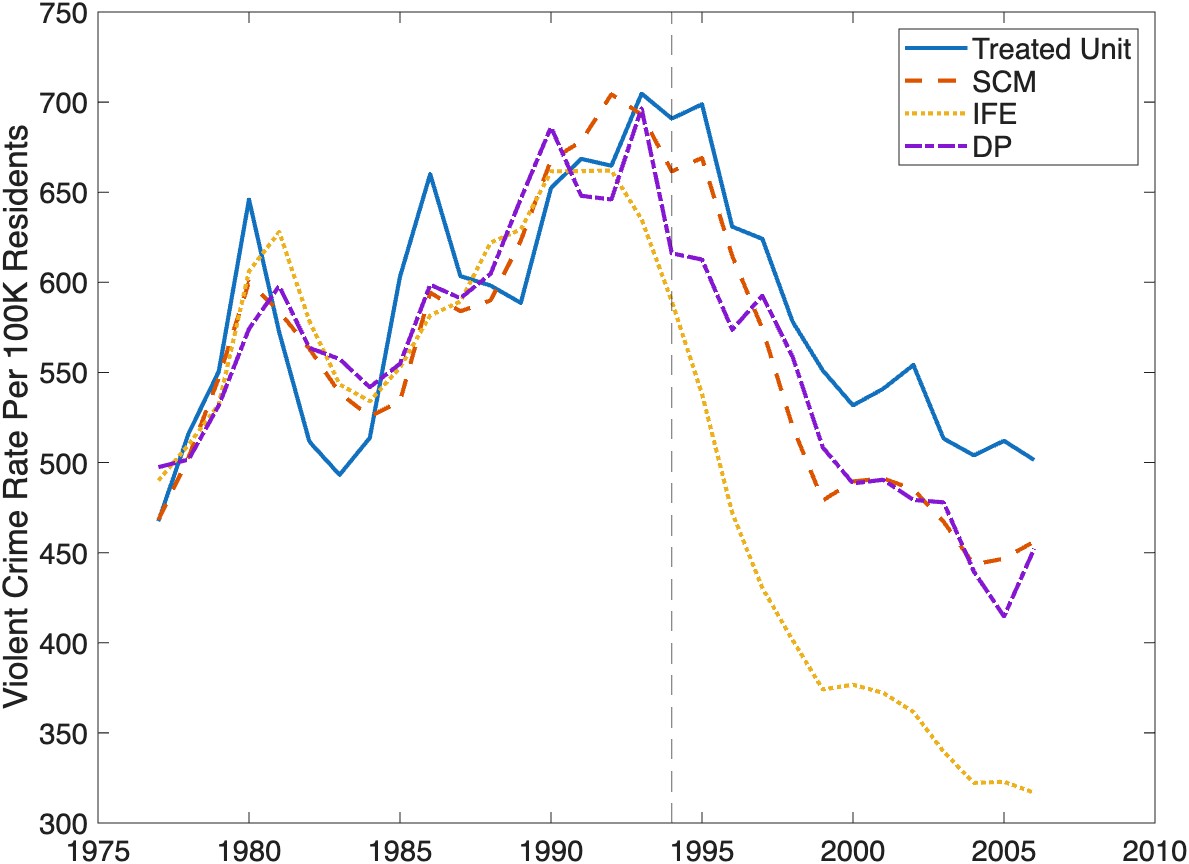}
    \caption{Treatment effect of RTC law of Arizona with different methods. The solid line is the actual violent crime rate. The dashed lines are the counterfactual outcomes with different methods. The vertical line denotes the beginning of the treatment effect}
    \label{figure3}
\end{figure}

For comparison, we evaluate two alternative methods, including interactive fixed effect (IFE) methods proposed by \mycite{701ec766-57f1-349b-b987-740b9cbdf712} and SCM proposed by \mycite{Abadie01062010}. Our analysis focuses on factor models with J=2. Since SCM requires all predictors are time-invariant, all explanatory covariates are averaged over pre-treatment periods, and  all pre-treatment outcomes are included\footnote{\mycite{https://doi.org/10.1002/pam.22206} highlighted a key limitation of SCM: the absence of clear guidelines for selecting predictor variables. They demonstrated that selectively including certain predictors—especially pre-treatment outcomes—to improve pre-treatment fit may lead to misleading inferences.}. As shown in \cref{figure3}, all methods follow a similar trend during the post-treatment periods. We can observe that the DP method closely aligns with SCM results, while the IFE method consistently has larger treatment effect estimates, diverging from the other methods. Lastly, since SCM in \mycite{Abadie01062010} relies on placebo tests for inference, our comparison in \cref{table4} focuses exclusively on the DP and IFE methods, evaluating the ATEL, standard error, and p-value. It is noteworthy that the ATEL estimated by the IFE method is substantially larger than that of the DP method, as observed in \cref{figure3}, resulting in a smaller p-value.

\begin{table}[H]
    \centering
    \begin{tabular}{c|c|c}
        J & ATEL & p-value \\
        \hline
        2 & 49.7665(15.0055 )& 0.0069 \\
        3 &  71.9248(12.3949 )& 0.0001 \\
        4 &  57.8420(13.3495 )& 0.0012 \\
        5 & 99.0401(13.7880 )& 0.0000 \end{tabular}
    \caption{Average treatment effect localization of RTC law of Arizona with different J. The number in the parenthesis is standard error }
    \label{table3}
\end{table}

\begin{table}[H]
    \centering
    \begin{tabular}{c|c|c}
        Method & ATEL & p-value \\
        \hline
        DP &  49.7665(15.0055 )& 0.0069 \\
        IFE & 153.5546(40.1636)&0.0024\\
    \end{tabular}
    \caption{Average treatment effect localization of RTC law on Arizona with DP and IFE. The number in the parenthesis is standard error }
    \label{table4}
\end{table}

 \section{Conclusion}\label{section7}
This paper introduced average treatment effect localization (ATEL) to capture localized, short-term impacts of policy interventions. We model untreated potential outcomes to flexibly account for both observed and unobserved covariates in a time-varying, nonparametric framework, which can be represented as a low-rank, time-varying factor structure via sieve approximation. By combining diversified projection for factor estimation with local linear smoothing for time-varying loadings, our method overcomes the limitations of conventional PCA-based approaches. We also develop the asymptotic distribution to facilitate inference on the ATEL. An empirical application on right-to-carry laws demonstrates the practical value of ATEL for timely policy evaluation.

\newpage
\printbibliography

\end{document}